\pdfoutput=1

\documentclass[11pt]{article}

\usepackage[preprint]{acl}

\usepackage{times}
\usepackage{latexsym}
\usepackage[T1]{fontenc}
\usepackage[utf8]{inputenc}
\usepackage{microtype}
\usepackage{inconsolata}
\usepackage{graphicx}

\usepackage{amsmath}
\usepackage{amssymb}
\usepackage{booktabs}
\usepackage{arydshln}
\usepackage{enumitem}
\usepackage{epstopdf}

\title{F5-TTS: A Fairytaler that Fakes Fluent and Faithful Speech \\ with Flow Matching}

\author{
  Yushen Chen$^1$, Zhikang Niu$^1$, Ziyang Ma$^1$, Keqi Deng$^2$\\ 
  \bf{Chunhui Wang$^3$, Jian Zhao$^3$, Kai Yu$^1$, Xie Chen$^1$\thanks{Corresponding author}} \\
  $^1$MoE Key Lab of Artificial Intelligence, X-LANCE Lab, Shanghai Jiao Tong University\\
  $^2$University of Cambridge $^3$Geely Automobile Research Institute (Ningbo) Company Ltd.\\
  \texttt{\{swivid,chenxie95\}@sjtu.edu.cn}
}

\begin{document}
\maketitle
\begin{abstract}
This paper introduces F5-TTS, a fully non-autoregressive text-to-speech system based on flow matching with Diffusion Transformer (DiT). 
Without requiring complex designs such as duration model, text encoder, and phoneme alignment, the text input is simply padded with filler tokens to the same length as input speech, and then the denoising is performed for speech generation, which was originally proved feasible by E2 TTS. 
However, the original design of E2 TTS makes it hard to follow due to its slow convergence and low robustness. 
To address these issues, we first model the input with ConvNeXt to refine the text representation, making it easy to align with the speech. 
We further propose an inference-time Sway Sampling strategy, which significantly improves our model's performance and efficiency. This sampling strategy for flow step can be easily applied to existing flow matching based models without retraining. 
Our design allows faster training and achieves an inference RTF of 0.15, which is greatly improved compared to state-of-the-art diffusion-based TTS models. 
Trained on a public 100K hours multilingual dataset, our F5-TTS exhibits highly natural and expressive zero-shot ability, seamless code-switching capability, and speed control efficiency. We have released all codes and checkpoints to promote community development, at \url{https://SWivid.github.io/F5-TTS/}. 
\end{abstract}

\section{Introduction}

\begin{figure*}[t]
\includegraphics[width=1\linewidth]{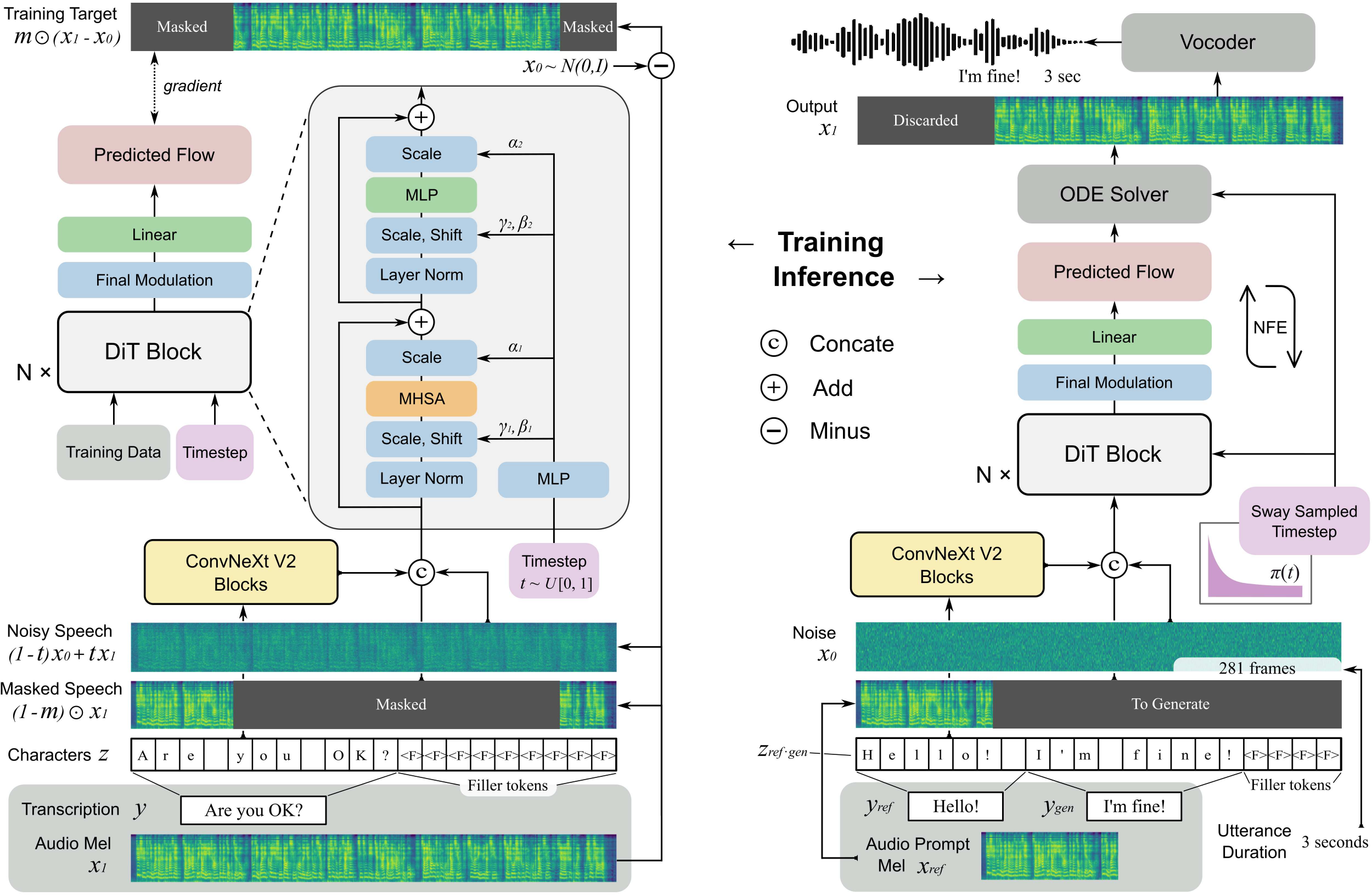}
\caption{An overview of F5-TTS training (left) and inference (right). The model is trained on the text-guided speech-infilling task and condition flow matching loss. The input text is converted to a character sequence, padded with filler tokens to the same length as input speech, and refined by ConvNeXt V2 blocks before concatenation with speech input. The inference leverages Sway Sampling for flow steps, with the model and an ODE solver to generate speech from sampled noise.}
\label{fig:overview}
\end{figure*}

Recent research in Text-to-Speech (TTS) has experienced great advancement~\citep{tacotron2,transtts,fastspeech2,glowtts,vits,gradtts,viola,ns1}. With a few seconds of audio prompt, current TTS models are able to synthesize speech for any given text and mimic the speaker of audio prompt~\citep{valle,vallex}. The synthesized speech can achieve high fidelity and naturalness that they are almost indistinguishable from human speech~\citep{ns2,ns3,valle2,voicebox}. 

While autoregressive (AR) based TTS models exhibit an intuitive way of consecutively predicting the next token(s) and have achieved promising zero-shot TTS capability, the inherent limitations of AR modeling require extra efforts addressing issues such as inference latency and exposure bias~\citep{ellav,vallt,valler,ralle,voicecraft}. Moreover, the quality of speech tokenizer is essential for AR models to achieve high-fidelity synthesis~\citep{soundstream,encodec,audiodec,hificodec,speechtokenizer,dmel,ndvq}. Thus, there have been studies exploring direct modeling in continuous space~\citep{ardittts,ar-wovq,melle} to enhance synthesized speech quality recently.

Although AR models demonstrate impressive zero-shot performance as they perform implicit duration modeling and can leverage diverse sampling strategies, non-autoregressive (NAR) models benefit from fast inference through parallel processing, and effectively balance synthesis quality and latency. Notably, diffusion models~\citep{ddpm,score} contribute most to the success of current NAR speech models~\citep{ns2,ns3}. In particular, Flow Matching with Optimal Transport path (FM-OT)~\citep{cfm-ot} is widely used in recent research fields not only text-to-speech~\citep{voicebox,voiceflow,matchatts,dittotts,e2tts} but also image generation~\citep{sd3} and music generation~\citep{fluxmusic}.

Unlike AR-based models, the alignment modeling between input text and synthesized speech is crucial and challenging for NAR-based models. While NaturalSpeech 3 \citep{ns3} and Voicebox \citep{voicebox} use frame-wise phoneme alignment; Matcha-TTS \citep{matchatts} adopts monotonic alignment search \citep{glowtts} and relies on a phoneme-level duration model; recent works find that introducing such rigid alignment between text and speech hinders the model from generating results with higher naturalness~\citep{e2tts,seedtts}. \\
\indent E3 TTS~\citep{e3tts} abandons phoneme-level duration and applies cross-attention on the input sequence but yields limited audio quality. DiTTo-TTS~\citep{dittotts} uses Diffusion Transformer (DiT)~\citep{dit} with cross-attention conditioned on encoded text from a pretrained language model. To further enhance alignment, it uses the pretrained language model to finetune the neural audio codec, infusing semantic information into the generated representations. In contrast, E2 TTS~\citep{e2tts}, based on Voicebox~\citep{voicebox}, adopts a simpler way, which removes the phoneme and duration predictor and directly uses characters padded with filler tokens to the length of mel spectrograms as input. This simple scheme also achieves very natural and realistic synthesized results. However, we found that robustness issues exist in E2 TTS for the text and speech alignment. Seed-TTS~\citep{seedtts} employs a similar strategy and achieves excellent results, though not elaborated in model details. In these ways of not explicitly modeling phoneme-level duration, models learn to assign the length of each word or phoneme according to the given total sequence length, resulting in improved prosody and rhythm. \\
\indent In this paper, we propose \textbf{F5-TTS}, a \textbf{F}airytaler that \textbf{F}akes \textbf{F}luent and \textbf{F}aithful speech with \textbf{F}low matching. Maintaining the simplicity of pipeline without phoneme alignment, duration predictor, text encoder, and semantically infused codec model, F5-TTS leverages the Diffusion Transformer with ConvNeXt V2~\citep{convnextv2} to better tackle text-speech alignment during in-context learning. We stress the deep entanglement of semantic and acoustic features in the E2 TTS model design, which has inherent problems and will pose alignment failure issues that could not simply be solved with re-ranking. With in-depth ablation studies, our proposed F5-TTS demonstrates stronger robustness, in generating more faithful speech to the text prompt, while maintaining comparable speaker similarity. Additionally, we introduce an inference-time sampling strategy for flow steps substantially improving naturalness, intelligibility, and speaker similarity of generation. This approach can be seamlessly integrated into existing flow matching based models without retraining.

\section{Preliminaries}

\subsection{Flow Matching}
\label{sec:fm}

The Flow Matching (FM) objective is to match a probability path $p_t$ from a simple distribution $\displaystyle p_0$, \textit{e.g.}, the standard normal distribution $p(x) = \mathcal{N}(x|0,I)$, to $\displaystyle p_1$ approximating the data distribution $q$. In short, the FM loss regresses the vector field $u_t$ with a neural network $v_t$ as 
\begin{equation}
\scalebox{0.9}{
$\displaystyle
\mathcal{L}_{\text{FM}}(\theta) = E_{t, p_t(x)} \left\| v_t(x) - u_t(x) \right\| ^2,
$}
\end{equation}
where $\theta$ parameterizes the neural network, $t\sim\mathcal{U}[0,1]$ and $x\sim p_t(x)$. $v_t$ is trained over the entire flow step and data range, ensuring it learns to handle the entire transformation process from the initial distribution to the target distribution.

As we have no prior knowledge of how to approximate $p_t$ and $u_t$, a conditional probability path $p_t(x|x_1) = \mathcal{N}(x\ |\ \mu_t(x_1),\sigma_t(x_1)^2I)$ is considered in actual training, and the Conditional Flow Matching (CFM) loss is proved to have identical gradients \textit{w.r.t.} $\theta$~\citep{cfm-ot}. $x_1$ is the random variable corresponding to training data. $\mu$ and $\sigma$ is the time-dependent mean and scalar standard deviation of Gaussian distribution.

Remember that the goal is to construct target distribution (data samples) from initial simple distribution, \textit{e.g.}, Gaussian noise. With the conditional form, the flow map $\psi_t(x)=\sigma_t(x_1)x+\mu_t(x_1)$ with $\mu_0(x_1)=0$ and $\sigma_0(x_1)=1$, $\mu_1(x_1)=x_1$ and $\sigma_1(x_1)=0$ is made to have all conditional probability paths converging to $p_0$ and $p_1$ at the start and end. The flow thus provides a vector field $d\psi_t(x_0)/dt = u_t(\psi_t(x_0)|x_1)$. Reparameterize $p_t(x|x_1)$ with $x_0$, we have 
\begin{equation}
\resizebox{1\linewidth}{!}{
$\displaystyle
\mathcal{L}_{\text{CFM}}(\theta) = E_{t, q(x_{1}), p(x_0)} \| v_{t}(\psi_t(x_0)) - \frac{d}{dt}\psi_t(x_0) \| ^2.
$}
\end{equation}
Further leveraging Optimal Transport (OT) form $\psi_t(x)=(1-t)x+tx_1$, the OT-CFM loss is
\begin{equation}
\resizebox{1\linewidth}{!}{
$\displaystyle
\mathcal{L}_{\text{CFM}}(\theta) = E_{t, q(x_{1}), p(x_0)} \| v_{t}((1-t)x_0+tx_1) - (x_1-x_0) \| ^2.
$}
\end{equation}
To view in a more general way~\citep{snrtheory}, if formulating the loss in terms of log signal-to-noise ratio (log-SNR) $\lambda$ instead of flow step $t$, and parameterizing to predict $x_0$ ($\epsilon$, commonly stated in diffusion model) instead of predict $x_1-x_0$, the CFM loss is equivalent to the v-prediction~\citep{vpredict} loss with cosine schedule.

For inference, given sampled noise $x_0$ from initial distribution $p_0$, flow step $t\in[0,1]$ and condition with respect to generation task, the ordinary differential equation (ODE) solver~\citep{torchdiffeq} is used to evaluate $\psi_1(x_0)$ the integration of $d\psi_t(x_0)/dt$ with $\psi_0(x_0)=x_0$. The number of function evaluations (NFE) is the times going through the neural network as we may provide multiple flow step values from 0 to 1 as input to approximate the integration. Higher NFE will produce more accurate results and certainly take more calculation time.

\subsection{Classifier-Free Guidance}
\label{sec:cfg}

Classifier Guidance (CG) is proposed by \citet{cg}, functions by adding the gradient of an additional classifier, while such an explicit way to condition the generation process may have several problems. Extra training of the classifier is required and the generation result is directly affected by the quality of the classifier. Adversarial attacks might also occur as the guidance is introduced through the way of updating the gradient. Thus deceptive images with imperceptible details to human eyes may be generated, which are not conditional.

Classifier-Free Guidance (CFG)~\citep{cfg} proposes to replace the explicit classifier with an implicit classifier without directly computing the explicit classifier and its gradient. The gradient of a classifier can be expressed as a combination of conditional generation probability and unconditional generation probability. By dropping the condition with a certain rate during training, and linear extrapolating the inference outputs with and without condition $c$, the final guided result is obtained. We could balance between fidelity and diversity of the generated samples with
\begin{equation}
\resizebox{1\linewidth}{!}{
$\displaystyle
v_{t,\text{CFG}} = v_{t}(\psi_t(x_0),c) + \alpha (v_{t}(\psi_t(x_0),c)-v_{t}(\psi_t(x_0)))
\label{eq:cfg}
$}
\end{equation}
in CFM case, where $\alpha$ is the CFG strength.\footnote{Note that the inference time will be doubled if CFG. Model $v_t$ will execute the forward process twice, once with condition, and once without.}

\section{Method}

This work aims to build a high-level text-to-speech synthesis system. We trained our model on the text-guided speech-infilling task \citep{a3t,voicebox}. Based on recent research \citep{dittotts,e2tts,e1tts}, it is promising to train without phoneme-level duration predictor and can achieve higher naturalness in zero-shot generation deprecating explicit phoneme-level alignment. We propose our advanced architecture with faster convergence and more robust generation. We also propose an inference-time flow step sampling strategy, which significantly improves our model's performance in faithfulness to reference text and speaker similarity.

\subsection{Pipeline}
\label{sec:pipeline}

\textbf{Training}\quad The infilling task is to predict a segment of speech given its surrounding audio and full text (for both surrounding transcription and the part to generate). For simplicity, we reuse the symbol $x$ to denote an audio sample and $y$ the corresponding transcript for a data pair $(x, y)$. As shown in Fig.\ref{fig:overview} (left), the acoustic input for training is an extracted mel spectrogram features $x_1\in \mathbb{R}^{F\times N}$ from the audio sample $x$, where $F$ is mel dimension and $N$ is the sequence length. In the scope of CFM, we pass in the model the noisy speech $(1-t)x_0+tx_1$ and the masked speech $(1-m)\odot x_1$, where $x_0$ denotes sampled Gaussian noise, $t$ is sampled flow step, and $m\in\{0,1\}^{F\times N}$ represents a binary temporal mask.

Following E2 TTS, we directly use alphabets and symbols for English. We opt for full pinyin to facilitate Chinese zero-shot generation. By breaking the raw text into such character sequence and padding it with filler tokens $\langle F \rangle$ to the same length as mel frames, we form an extended sequence $z$ with $c_i$ denoting the $i$-th character:
\begin{equation}
z = (c_1, c_2, \ldots, c_M, \underbrace{\langle F \rangle, \ldots, \langle F \rangle}_{(N-M)\text{ times}}).
\end{equation}
The model is trained to reconstruct $m\odot x_1$ with $(1-m)\odot x_1$ and $z$, which equals to learn the target distribution $p_1$ in form of $P(m\odot x_1|(1-m)\odot x_1,z)$ approximating real data distribution $q$.

\noindent
\textbf{Inference}\quad To generate a speech with the desired content, we have the audio prompt's mel spectrogram features $x_{ref}$, its transcription $y_{ref}$, and a text prompt $y_{gen}$. Audio prompt serves to provide speaker characteristics and text prompt is to guide the content of generated speech.

The sequence length $N$, or duration, has now become a pivotal factor that necessitates informing the model of the desired length for sample generation. One could train a separate model to predict and deliver the duration based on $x_{ref}$, $y_{ref}$ and $y_{gen}$. Here we simply estimate the duration based on the ratio of the number of characters in $y_{gen}$ and $y_{ref}$. We assume that the sum-up length of characters is no longer than mel length, thus padding with filler tokens is done as during training.

To sample from the learned distribution, the converted mel features $x_{ref}$, along with concatenated and extended character sequence $z_{ref\cdot gen}$ serve as the condition in Eq.\ref{eq:cfg}. We have
\begin{equation}
v_t(\psi_t(x_0),c) = v_t((1-t)x_0+tx_1|x_{ref}, z_{ref\cdot gen}), 
\end{equation}
See from Fig.\ref{fig:overview} (right), we start from a sampled noise $x_0$, and what we want is the other end of flow $x_1$. Thus we use the ODE solver to gradually integrate from $\psi_0(x_0)=x_0$ to $\psi_1(x_0)=x_1$, given $d\psi_t(x_0)/dt=v_t(\psi_t(x_0),x_{ref}, z_{ref\cdot gen})$. During inference, the flow steps are provided in an ordered way, $e.g.$, uniformly sampled a certain number from 0 to 1 according to the NFE setting.
After getting the generated mel with model $v_t$ and ODE solver, we discard the part of $x_{ref}$. Then we leverage a vocoder to convert the mel back to waveform.

\subsection{F5-TTS}
\label{sec:f5tts}

E2 TTS directly concatenates the padded character sequence with input speech sequence, deeply entangling semantic and acoustic features with a large length gap of effective information, which is the underlying cause of hard training and poses several problems in a zero-shot scenario (Sec.\ref{sec:modelarchitecture}). To alleviate the problem of slow convergence and low robustness, we propose F5-TTS which accelerates training and inference and shows a strong robustness in generation. Also, an inference-time Sway Sampling is introduced, which allows inference faster (using less NFE) while maintaining performance. This sampling way of flow step can be directly applied to other CFM-based models without retraining.

\noindent
\textbf{Model}\quad As shown in Fig.\ref{fig:overview}, we use latent Diffusion Transformer (DiT)~\citep{dit} as backbone. To be specific, we use DiT blocks with zero-initialized adaptive Layer Norm (adaLN-zero). To enhance the model's alignment ability, we also leverage ConvNeXt V2 blocks~\citep{convnextv2}. Its predecessor ConvNeXt V1~\citep{convnext} is used in many works and shows a strong temporal modeling capability in speech domain tasks~\citep{vocos,convnexttts}. 

As described in Sec.\ref{sec:pipeline}, the model input is character sequence, noisy speech, and masked speech. Before concatenation in the feature dimension, the character sequence first goes through ConvNeXt blocks. Experiments have shown that this way of providing individual modeling space allows text input to better prepare itself before later in-context learning. Unlike the phoneme-level force alignment done in Voicebox, a rigid boundary for text is not explicitly introduced. The semantic and acoustic features are jointly learned with the entire model. Not directly feeding the model with inputs of significant length gap as E2 TTS does, the proposed text refinement mitigates the impact of using inputs with mismatched effective information lengths, despite equal physical length in magnitude as E2 TTS.

The flow step $t$ for CFM is provided as the condition of adaLN-zero rather than appended to the concatenated input sequence in Voicebox. We found that an additional mean pooled token of text sequence for adaLN condition is not essential for the TTS task, maybe because the TTS task requires more rigorously guided results and the mean pooled text token is more coarse.

We adopt some position embedding settings in Voicebox. The flow step is embedded with a sinusoidal position. The concatenated input sequence is added with a convolutional position embedding. We apply a rotary position embedding (RoPE)~\citep{rope} for self-attention rather than symmetric bi-directional ALiBi bias~\citep{alibi}. And for extended character sequence $z$, we also add it with an absolute sinusoidal position embedding before feeding it into ConvNeXt blocks.

Compared with Voicebox and E2 TTS, we abandoned the U-Net~\citep{unet} style skip connection structure and switched to using DiT with adaLN-zero. Without a phoneme-level duration predictor and explicit alignment process, and nor with extra text encoder and semantically infused neural codec model in DiTTo-TTS, we give the text input a little freedom (individual modeling space) to let it prepare itself before concatenation and in-context learning with speech input.

\noindent
\textbf{Sampling}\quad As stated in Sec.\ref{sec:fm}, the CFM could be viewed as v-prediction with a cosine schedule. For image synthesis, \citet{sd3} propose to further schedule the flow step with a single-peak logit-normal~\citep{lognorm} sampling, in order to give more weight to intermediate flow steps by sampling them more frequently. We speculate that such sampling distributes the model's learning difficulty more evenly over different flow step $t\in[0,1]$.

In contrast, we train our model with traditional uniformly sampled flow step $t\sim \mathcal{U}[0,1]$ but apply a non-uniform sampling during inference. In specific, we define a \textbf{Sway Sampling} function as
\begin{equation}
f_{sway}(u;s) = u+s\cdot(\cos(\frac{\pi}{2}u)-1+u),
\end{equation}
which is monotonic with coefficient $s\in[-1,\frac{2}{\pi-2}]$. We first sample $u\sim \mathcal{U}[0,1]$, then apply this function to obtain sway sampled flow step $t$. With $s<0$, the sampling is sway to left; with $s>0$, the sampling is sway to right; and $s=0$ case equals to uniform sampling. Fig.\ref{fig:swaysampling} shows the probability density function of Sway Sampling on flow step $t$. 

Conceptually, CFM models focus more on sketching the contours of speech in the early stage ($t\to0$) from pure noise and later focus more on the embellishment of fine-grained details. Therefore, the alignment between speech and text will be determined based on the first few generated results. With a scale parameter $s<0$, we make model inference more with smaller $t$, thus providing the ODE solver with more startup information to evaluate more precisely in initial integration steps.

\section{Experimental Setup}

\textbf{Datasets}\quad We utilize the in-the-wild multilingual speech dataset Emilia~\citep{emilia} to train our base models. After simply filtering out transcription failure and misclassified language speech, we retain approximately 95K hours of English and Chinese data. We also trained small models for ablation study and architecture search on WenetSpeech4TTS~\citep{wenet4tts} Premium subset, consisting of a 945 hours Mandarin corpus. Base model configurations are introduced below, and small model configurations are in Appendix \ref{appx:smallmodels}. Three test sets are adopted for evaluation, which are LibriSpeech-PC \textit{test-clean}~\citep{librispeechpc}, Seed-TTS \textit{test-en}~\citep{seedtts} with 1088 samples from Common Voice~\citep{commonvoice}, and Seed-TTS \textit{test-zh} with 2020 samples from DiDiSpeech~\citep{didispeech}\footnote{\url{https://github.com/BytedanceSpeech/seed-tts-eval}}. Most of the previous English-only models are evaluated on different subsets of LibriSpeech \textit{test-clean} while the used prompt list is not released, which makes fair comparison difficult. Thus we build and release a 4-to-10-second LibriSpeech-PC subset with 1127 samples to facilitate community comparisons.

\noindent
\textbf{Training}\quad Our base models are trained to 1.2M updates with a batch size of 307,200 audio frames (0.91 hours), for over one week on 8 NVIDIA A100 80G GPUs. The AdamW optimizer~\citep{adamw} is used with a peak learning rate of 7.5e-5, linearly warmed up for 20K updates, and linearly decays over the rest of the training. We set 1 for the max gradient norm clip.
The F5-TTS base model has 22 layers, 16 attention heads, 1024/2048 embedding/feed-forward network (FFN) dimension for DiT; and 4 layers, 512/1024 embedding/FFN dimension for ConvNeXt V2; in total 335.8M parameters. The reproduced E2 TTS, a 333.2M flat U-Net equipped Transformer, has 24 layers, 16 attention heads, and 1024/4096 embedding/FFN dimension. Both models use RoPE as mentioned in Sec.\ref{sec:f5tts}, a dropout rate of 0.1 for attention and FFN, the same convolutional position embedding as in Voicebox\citep{voicebox}.

We directly use alphabets and symbols for English, use jieba\footnote{\url{https://github.com/fxsjy/jieba}} and pypinyin\footnote{\url{https://github.com/mozillazg/python-pinyin}} to process raw Chinese characters to full pinyins. The character embedding vocabulary size is 2546, counting in the special filler token and all other language characters exist in the Emilia dataset as there are many code-switched sentences. For audio samples we use 100-dimensional log mel-filterbank features with 24 kHz sampling rate and hop length 256. A random 70\% to 100\% of mel frames is masked for infilling task training. For CFG (Sec.\ref{sec:cfg}) training, first the masked speech input is dropped with a rate of 0.3, then the masked speech again but with text input together is dropped with a rate of 0.2 \citep{voicebox}. We assume that the two-stage control of CFG training may have the model learn more with text alignment.

\noindent
\textbf{Inference}\quad The inference process is mainly elaborated in Sec.\ref{sec:pipeline}. We use the Exponential Moving Averaged (EMA)~\citep{ema} weights for inference, and the Euler ODE solver for F5-TTS (midpoint for E2 TTS as described in \citet{e2tts}). We use the pretrained vocoder Vocos~\citep{vocos} to convert generated log mel spectrograms to audio signals.

\noindent
\textbf{Baselines}\quad We compare our models with leading TTS systems including, (mainly)
autoregressive models: VALL-E 2~\citep{valle2}, MELLE~\citep{melle}, FireRedTTS~\citep{fireredtts} and CosyVoice~\citep{cosyvoice}; 
non-autoregressive models: Voicebox~\citep{voicebox}, NaturalSpeech 3~\citep{ns3}, DiTTo-TTS~\citep{dittotts}, MaskGCT~\citep{maskgct}, Seed-TTS$_{DiT}$~\citep{seedtts} and our reproduced E2 TTS~\citep{e2tts}. Details of compared models see Appendix~\ref{appx:baselines}.

\noindent
\textbf{Metrics}\quad We measure the performances under \textit{cross-sentence} task \citep{valle, voicebox}. 
We report Word Error Rate (WER) and speaker Similarity between generated and the original target speeches (SIM-o \citep{voicebox}) for objective evaluation. For WER, we employ Whisper-large-v3~\citep{whisper} to transcribe English and Paraformer-zh~\citep{funasr} for Chinese, following~\citep{seedtts}. For SIM-o, we use a WavLM-large-based~\citep{wavlm} speaker verification model to extract speaker embeddings for calculating the cosine similarity of synthesized and ground truth speeches. 
We use Comparative Mean Opinion Scores (CMOS) and Similarity Mean Opinion Scores (SMOS) for subjective evaluation. 
Details of subjective evaluations can be found in Appendix \ref{appx:subj_eval}.

\section{Experimental Results}
\label{sec:expr_results}

Tab.\ref{tab:librispeech-test} and \ref{tab:seedtts-test} show the main results of objective and subjective evaluations. We report the average score of three random seed generation results with our model and open-sourced baselines. We use by default a CFG strength of 2 and a Sway Sampling coefficient of $-1$ for our F5-TTS.

\begin{table}[t]
\begin{center}
\resizebox{1\linewidth}{!}{
\begin{tabular}{llllll}
\toprule
\multicolumn{6}{l}{\bf Model \hspace{4.0em} \#Param. \hspace{0.4em} \#Data \hspace{0.3em} WER(\%)↓ \ \bf SIM-o↑ { } \bf RTF↓} \\
\midrule
\multicolumn{6}{c}{\textbf{LibriSpeech \textit{test-clean}}} \\
\midrule
\multicolumn{3}{l}{Ground Truth (\textit{2.2 hours subset})}           &2.2           &0.754          &{ }{ }- \\
\cdashline{1-6}\noalign{\vskip\belowrulesep}
\multicolumn{2}{l}{VALL-E 2          \hspace{4.3em} -}    &{ }{ }50K EN      &2.44          &0.643          &0.732 \\
\multicolumn{2}{l}{MELLE             \hspace{5.2em} -}    &{ }{ }50K EN      &2.10          &0.625          &0.549 \\
\multicolumn{2}{l}{MELLE-\textit{R2} \hspace{3.75em} -}   &{ }{ }50K EN      &2.14          &0.608          &0.276 \\
\multicolumn{2}{l}{Voicebox  \hspace{3.8em}  330M}        &{ }{ }60K EN      &\textbf{1.9}  &\textbf{0.662} &0.64  \\
\multicolumn{2}{l}{DiTTo-TTS \hspace{2.8em}  740M}        &{ }{ }55K EN      &2.56          &0.627          &\textbf{0.162} \\
\midrule
\multicolumn{3}{l}{Ground Truth (\textit{40 samples subset})}          &1.94          &0.68           &{ }{ }-  \\
\cdashline{1-6}\noalign{\vskip\belowrulesep}
\multicolumn{2}{l}{Voicebox \hspace{3.8em} 330M}          &{ }{ }60K EN      &2.03          &0.64           &0.64  \\
\multicolumn{2}{l}{NaturalSpeech 3 \hspace{0.9em}  500M}  &{ }{ }60K EN      &\textbf{1.94} &0.67           &0.296 \\
\multicolumn{2}{l}{MaskGCT    \hspace{2.8em} 1048M}       &100K Multi. &2.634         &\textbf{0.687} &{ }{ }-  \\
\midrule
\multicolumn{6}{c}{\textbf{LibriSpeech-PC \textit{test-clean}}} \\
\midrule
\multicolumn{3}{l}{Ground Truth (\textit{1127 samples 2 hrs})}         &2.23          &0.69           &{ }{ }-  \\
\multicolumn{3}{l}{Vocoder Resynthesized}                              &2.32          &0.66           &{ }{ }-  \\
\cdashline{1-6}\noalign{\vskip\belowrulesep}
\multicolumn{2}{l}{CosyVoice  \hspace{2.6em} $\sim$300M}  &170K Multi. &3.59          &0.66           &0.92\\
\multicolumn{2}{l}{FireRedTTS \hspace{1.9em} $\sim$580M}  &248K Multi. &2.69          &0.47           &0.84\\
\multicolumn{2}{l}{E2 TTS (32 NFE) \hspace{0.5em} 333M}   &100K Multi. &2.95          &\textbf{0.69}  &0.68\\
\cdashline{1-6}\noalign{\vskip\belowrulesep}
\multicolumn{2}{l}{F5-TTS (16 NFE) \hspace{0.5em} 336M}   &100K Multi. &2.53          &0.66           &\textbf{0.15}\\
\multicolumn{2}{l}{F5-TTS (32 NFE) \hspace{0.5em} 336M}   &100K Multi. &\textbf{2.42} &0.66           &0.31\\
\bottomrule
\end{tabular}
}
\end{center}
\caption{Comparison results on LibriSpeech(-PC) \textit{test-clean}. The Real-Time Factor (RTF) is computed with the inference time of 10s speech on NVIDIA RTX 3090. \#Param. stands for the number of learnable parameters and \#Data refers to the used training dataset in hours.}
\label{tab:librispeech-test}
\end{table}

F5-TTS achieves a WER of 2.42 on LibriSpeech-PC \textit{test-clean} with 32 NFE, demonstrating its robustness in zero-shot generation. Inference with 16 NFE, F5-TTS gains an RTF of 0.15 while still supporting high-quality generation with a WER of 2.53. The reproduced E2 TTS shows an excellent speaker similarity (SIM) but much worse WER in the zero-shot scenario, indicating the inherent deficiency of alignment robustness. 

From the evaluation results on the Seed-TTS test sets, F5-TTS behaves similarly with a close WER to ground truth and comparable SIM scores. It produces smooth and fluent speech in zero-shot generation with a CMOS of 0.31 (0.21) and SMOS of 3.89 (3.83) on Seed-TTS \textit{test-en} (\textit{test-zh}), and surpasses some baseline models trained with larger scales. As stated in Sec.\ref{sec:pipeline}, we simply estimate duration based on the ratio of the audio prompt's transcript length and the text prompt length, rather than relying on an extra duration predictor. It is also worth mentioning that Seed-TTS with the best result is trained with orders of larger model size and dataset (several million hours) than ours.

A robustness test on ELLA-V \citep{ellav} hard sentences is further included in Appendix \ref{appx:ellavhardtest}. The ablation of vocoders and additional evaluation with a non-PC test set are in Appendix \ref{appx:compr_vocoder}. An analysis of training stability with varying data scales is in Appendix \ref{appx:compr_data_scale}.

\begin{table}[t]
\begin{center}
\resizebox{1\linewidth}{!}{
\begin{tabular}{lllcc}
\toprule
\multicolumn{5}{l}{\bf Model \hspace{5.5em} \bf WER(\%)↓ { } \bf SIM-o↑ { }{ } \bf CMOS↑ { } \bf SMOS↑} \\
\midrule
\multicolumn{5}{c}{\textbf{Seed-TTS \textit{test-en}}} \\
\midrule
Ground Truth                 &\ \ 2.06             &\ 0.73               &{\,\,}0.00             &3.91 \\
Vocoder Resynthesized        &\ \ 2.09             &\ 0.70               &-                      &- \\
\cdashline{1-5}\noalign{\vskip\belowrulesep}
CosyVoice                    &\ \ 3.39             &\ 0.64               &{\,\,}0.02             &3.64 \\
FireRedTTS                   &\ \ 3.82             &\ 0.46               &-1.46                  &2.94 \\
MaskGCT                      &\ \ 2.623*           &\ \underline{0.717}* &-                      &- \\
Seed-TTS$_{DiT}$             &\ \ \textbf{1.733}*  &\ \textbf{0.790}*    &-                      &- \\
E2 TTS (32 NFE)              &\ \ 2.19             &\ 0.71               &{\,\,}0.06             &\underline{3.81} \\
\cdashline{1-5}\noalign{\vskip\belowrulesep}
F5-TTS (16 NFE)              &\ \ 1.89             &\ 0.67               &{\,\,}\underline{0.16} &3.79 \\
F5-TTS (32 NFE)              &\ \ \underline{1.83} &\ 0.67               &{\,\,}\textbf{0.31}    &\textbf{3.89} \\
\midrule
\multicolumn{5}{c}{\textbf{Seed-TTS \textit{test-zh}}} \\
\midrule
Ground Truth                 &\ \ 1.26             &\ 0.76               &{\,\,}0.00             &3.72 \\
Vocoder Resynthesized        &\ \ 1.27             &\ 0.72               &-                      &- \\
\cdashline{1-5}\noalign{\vskip\belowrulesep}
CosyVoice                    &\ \ 3.10             &\ 0.75               &-0.06                  &3.54 \\
FireRedTTS                   &\ \ \underline{1.51} &\ 0.63               &-0.49                  &3.28 \\
MaskGCT                      &\ \ 2.273*           &\ \underline{0.774}* &-                      &- \\
Seed-TTS$_{DiT}$             &\ \ \textbf{1.178}*  &\ \textbf{0.809}*    &-                      &- \\
E2 TTS (32 NFE)              &\ \ 1.97             &\ 0.73               &-0.04                  &3.44 \\
\cdashline{1-5}\noalign{\vskip\belowrulesep}
F5-TTS (16 NFE)              &\ \ 1.74             &\ 0.75               &{\,\,}\underline{0.02} &\underline{3.72} \\
F5-TTS (32 NFE)              &\ \ 1.56             &\ 0.76               &{\,\,}\textbf{0.21}    &\textbf{3.83} \\
\bottomrule
\end{tabular}
}
\end{center}
\caption{Results on two test sets, Seed-TTS \textit{test-en} and \textit{test-zh}. The boldface indicates the best result, the underline denotes the second best, and * denotes scores reported in baseline papers.}
\label{tab:seedtts-test}
\end{table}

\subsection{Ablation of Model Architecture}
\label{sec:modelarchitecture}

To clarify our F5-TTS's efficiency and stress the limitation of E2 TTS. We conduct in-depth ablation studies. We trained small models (all around 155M parameters) to 800K updates on the WenetSpeech4TTS Premium 945 hours Mandarin dataset with half the batch size compared to base models. Configuration details see Appendix \ref{appx:smallmodels}.

\begin{figure}[ht]
\begin{center}
\includegraphics[width=1\linewidth]{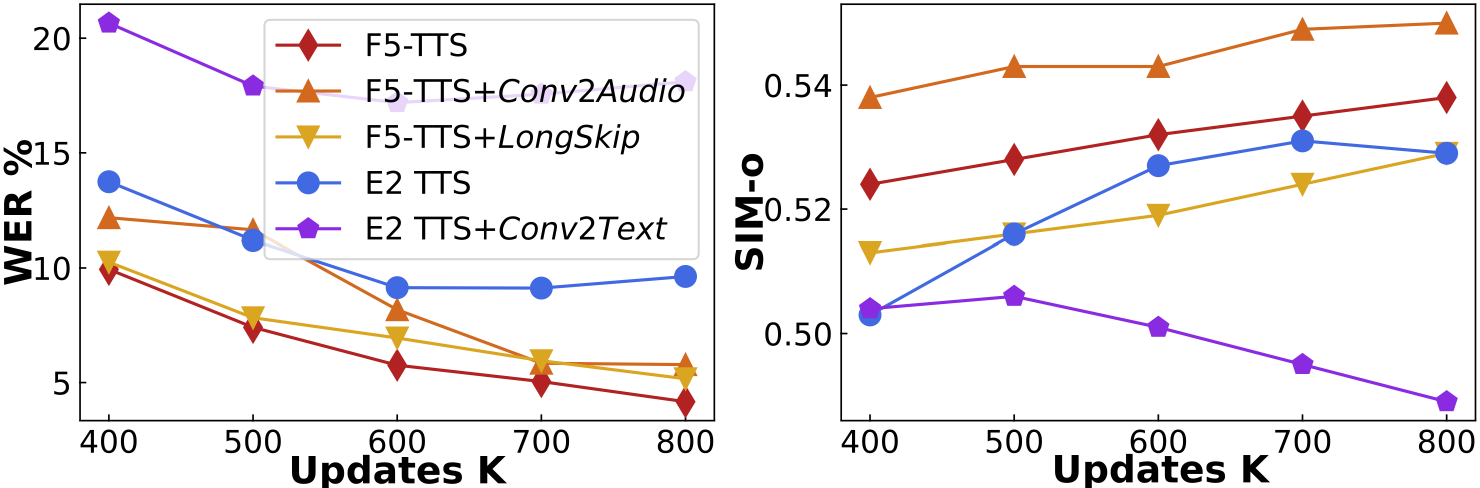}
\end{center}
\caption{Seed-TTS \textit{test-zh} evaluation results of 155M small models trained with WenetSpeech4TTS Premium a 945 hours Mandarin Corpus.}
\label{fig:modelarchitecture}
\end{figure}

Fig.\ref{fig:modelarchitecture} shows the overall trend of productive small models' WER and SIM scores evaluated on Seed-TTS \textit{test-zh}. F5-TTS (32 NFE \textit{w/o} SS) achieves a WER of 4.17 and a SIM of 0.54 at 800K updates, while E2 TTS is 9.63 and 0.53. 
We were first faced with unsatisfactory results with reproduced E2 TTS in Mandarin. On the one hand, E2 TTS converges slowly; on the other hand, we found that E2 TTS consistently failed (unable to solve with re-ranking) on 7\% test samples (WER$>$50\%) all along the training process, speculated of larger distribution gap with train set.
To disclose the possible reasons for E2 TTS's deficiency, we investigate the models' behaviors with different inputs. See from Tab.\ref{tab:ablationinput} in Appendix~\ref{appx:ablationinput}, by dropping the audio prompt, and synthesizing speech solely with the text prompt, E2 TTS is free of failures (F5-TTS also benefits but because of more standard output that is easier to be recognized by the ASR model). This phenomenon implied a deep entanglement of semantic and acoustic features within E2 TTS's model design, and it greatly hinders real-world application as the failed generation cannot be solved with re-ranking. Supervised fine-tuning facing out-of-domain data or a tremendous pretraining scale under the slow convergence speed is mandatory for E2 TTS, which is inconvenient for industrial deployment and a crushing burden for individuals.

From Tab.\ref{tab:smallmodels} GFLOPs, structures without many skip connections natively allow faster training and inference. However, pure adaLN DiT (F5-TTS\textit{$-$Conv2Text}) failed to learn alignment given simply padded character sequences, while MMDiT \citep{sd3} learned fast and collapsed fast, resulting in severe repeated utterance with wild timbre and prosody. We assume that the pure MMDiT structure is far too flexible for TTS task that requires faithful generation following guidance. Thus we focus on enhancing the modeling ability of DiT. Based on the concept of refining the input text representation to better align with speech modality, and keep the simplicity of system design, F5-TTS is proven effective. F5-TTS easily handles zero-shot generation, showing stronger robustness.

We further ablate with adding the same branch for input speech (F5-TTS\textit{$+$Conv2Audio}), and also conduct experiments to figure out whether the long skip connection and the refinement of input text are beneficial to the counterpart backbone, \textit{i.e.} F5-TTS and E2 TTS, named F5-TTS\textit{$+$LongSkip} and E2 TTS\textit{$+$Conv2Text} respectively.
From Fig.\ref{fig:modelarchitecture}, F5-TTS\textit{$+$Conv2Audio} trades much alignment robustness (+1.61 WER) with a slightly higher speaker similarity (+0.01 SIM). The long skip connection structure can not simply fit into DiT to improve speaker similarity, while the ConvNeXt for input text refinement can not directly apply to the flat U-Net Transformer to improve WER as well, both showing significant degradation of performance.

\subsection{Ablation of Sway Sampling}
\label{sec:swaysampling}

\begin{figure}[t]
\begin{center}
\includegraphics[width=1\linewidth]{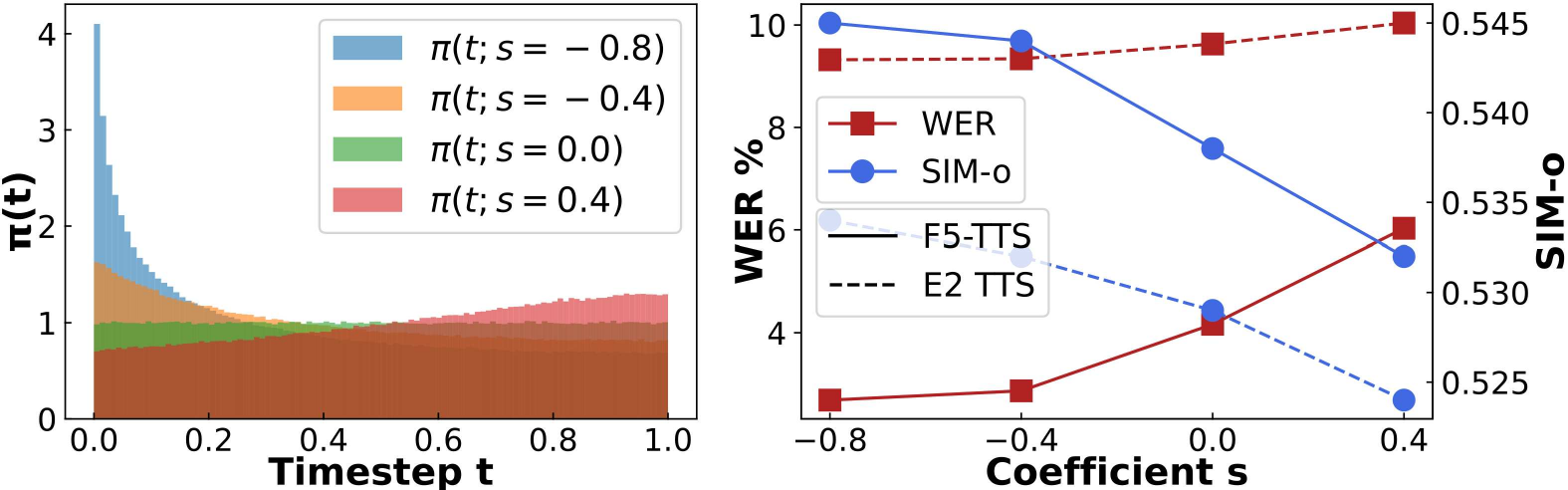}
\end{center}
\caption{The probability density function of Sway Sampling with different coefficient $s$, and small models' corresponding performance on Seed-TTS \textit{test-zh}.}
\label{fig:swaysampling}
\end{figure}

It is clear from Fig.\ref{fig:swaysampling} that a Sway Sampling with more negative $s$ value further improves performance. Appendix \ref{appx:swaysampling} with massive ablation results on base models, provides more evidence of the effectiveness of the proposed strategy.

To be more concrete and intuitive, we conduct a "leak and override" experiment. We first replace the Gaussian noise input $x_0$ at inference time with a ground-truth-information-leaked input $(1-t')x_0+t'x_{ref}'$, where $t'=0.1$ and $x_{ref}'$ is a duplicate of the audio prompt mel features. Then, we provide a text prompt different from the duplicated audio transcript and let the model continue the subsequent inference (skip the flow steps before $t'$). The model succeeds in overriding leaked utterances and producing speech following the text prompt if Sway Sampling is used, and fails without. Uniformly sampled flow steps will have the model producing speech dominated by leaked information, speaking the duplicated audio prompt's context. Similarly, a leaked timbre can be overridden with another speaker's utterance as an audio prompt, leveraging Sway Sampling. 

The experiment result is a shred of strong evidence proving that the early flow steps are crucial for sketching the silhouette of target speech based on given prompts faithfully, the later steps focus more on formed intermediate noisy output, where our sway-to-left sampling ($s<0$) finds the profitable niche and takes advantage of it. We emphasize that our inference-time Sway Sampling can be easily applied to existing CFM-based models without retraining. And we will work in the future to combine it with training-time noise schedulers and distillation techniques to further boost efficiency.

\section{Conclusion}
This work introduces F5-TTS, a fully non-autoregressive text-to-speech system based on flow matching with diffusion transformer (DiT). 
With a tidy pipeline, literally text in and speech out, F5-TTS achieves state-of-the-art zero-shot ability compared to existing works trained on industry-scale data. We adopt ConvNeXt for text modeling and propose the test-time Sway Sampling strategy to further improve the robustness of speech generation and inference efficiency.
Our design allows faster training and inference, by achieving a test-time RTF of 0.15, which is competitive with other heavily optimized TTS models of similar performance. 
We will open-source our code, and models, to enhance transparency and facilitate reproducible research in this area.

\section*{Limitations}

There are two limitations to this work. 
First, although F5-TTS accelerates the training and inference while maintaining the simplicity of the system through better structure design, the mel spectrogram sequence length is still much longer than the text modality. Therefore, research and employment of a more efficient and hopefully universal continuous representation compatible with highly expressive speech synthesis remains a critical direction and can further improve efficiency and performance.
Second, although F5-TTS has great zero-shot generation ability and can deeply mimic the reference audio, it lacks fine-grained control of paralinguistic details, \textit{e.g.} emotion, which is of great research and practical application value.

\section*{Ethics Statements}
This work is purely a research project. F5-TTS is trained on large-scale public multilingual speech data and could synthesize speech of high naturalness and speaker similarity. Given the potential risks in the misuse of the model, such as spoofing voice identification, it should be imperative to implement watermarks and develop a detection model to identify audio outputs.



\bibliography{custom}

\begin{thebibliography}{76}
\providecommand{\natexlab}[1]{#1}

\bibitem[{Anastassiou et~al.(2024)Anastassiou, Chen, Chen, Chen et~al.}]{seedtts}
Philip Anastassiou, Jiawei Chen, Jitong Chen, Yuanzhe Chen, et~al. 2024.
\newblock {Seed-TTS}: A family of high-quality versatile speech generation models.
\newblock \emph{arXiv preprint arXiv:2406.02430}.

\bibitem[{Ardila et~al.(2019)Ardila, Branson, Davis, Henretty, Kohler, Meyer, Morais, Saunders, Tyers, and Weber}]{commonvoice}
Rosana Ardila, Megan Branson, Kelly Davis, Michael Henretty, Michael Kohler, Josh Meyer, Reuben Morais, Lindsay Saunders, Francis~M Tyers, and Gregor Weber. 2019.
\newblock Common voice: A massively-multilingual speech corpus.
\newblock \emph{arXiv preprint arXiv:1912.06670}.

\bibitem[{Atchison and Shen(1980)}]{lognorm}
Jhon Atchison and Sheng~M Shen. 1980.
\newblock Logistic-normal distributions: Some properties and uses.
\newblock \emph{Biometrika}, 67(2):261--272.

\bibitem[{Bai et~al.(2024)Bai, Likhomanenko, Zhang, Gu, Aldeneh, and Jaitly}]{dmel}
He~Bai, Tatiana Likhomanenko, Ruixiang Zhang, Zijin Gu, Zakaria Aldeneh, and Navdeep Jaitly. 2024.
\newblock {dMel}: Speech tokenization made simple.
\newblock \emph{arXiv preprint arXiv:2407.15835}.

\bibitem[{Bai et~al.(2022)Bai, Zheng, Chen, Ma, Li, and Huang}]{a3t}
He~Bai, Renjie Zheng, Junkun Chen, Mingbo Ma, Xintong Li, and Liang Huang. 2022.
\newblock A3t: Alignment-aware acoustic and text pretraining for speech synthesis and editing.
\newblock In \emph{International Conference on Machine Learning}, pages 1399--1411. PMLR.

\bibitem[{Chen(2018)}]{torchdiffeq}
Ricky T.~Q. Chen. 2018.
\newblock \href {https://github.com/rtqichen/torchdiffeq} {torchdiffeq}.

\bibitem[{Chen et~al.(2024)Chen, Liu, Zhou, Liu, Tan, Li, Zhao, Qian, and Wei}]{valle2}
Sanyuan Chen, Shujie Liu, Long Zhou, Yanqing Liu, Xu~Tan, Jinyu Li, Sheng Zhao, Yao Qian, and Furu Wei. 2024.
\newblock {VALL-E 2}: Neural codec language models are human parity zero-shot text to speech synthesizers.
\newblock \emph{arXiv preprint arXiv:2406.05370}.

\bibitem[{Chen et~al.(2022)Chen, Chen, Wu, Qian, Wang, Liu, Qian, and Zeng}]{wavlm}
Zhengyang Chen, Sanyuan Chen, Yu~Wu, Yao Qian, Chengyi Wang, Shujie Liu, Yanmin Qian, and Michael Zeng. 2022.
\newblock Large-scale self-supervised speech representation learning for automatic speaker verification.
\newblock In \emph{Proc. ICASSP}, pages 6147--6151. IEEE.

\bibitem[{D{\'e}fossez et~al.(2022)D{\'e}fossez, Copet, Synnaeve, and Adi}]{encodec}
Alexandre D{\'e}fossez, Jade Copet, Gabriel Synnaeve, and Yossi Adi. 2022.
\newblock High fidelity neural audio compression.
\newblock \emph{arXiv preprint arXiv:2210.13438}.

\bibitem[{Dhariwal and Nichol(2021)}]{cg}
Prafulla Dhariwal and Alexander Nichol. 2021.
\newblock Diffusion models beat gans on image synthesis.
\newblock \emph{Advances in neural information processing systems}, 34:8780--8794.

\bibitem[{Du et~al.(2024{\natexlab{a}})Du, Guo, Wang, Yang, Niu, Wang, Zhang, Chen, and Yu}]{vallt}
Chenpeng Du, Yiwei Guo, Hankun Wang, Yifan Yang, Zhikang Niu, Shuai Wang, Hui Zhang, Xie Chen, and Kai Yu. 2024{\natexlab{a}}.
\newblock {VALL-T}: Decoder-only generative transducer for robust and decoding-controllable text-to-speech.
\newblock \emph{arXiv preprint arXiv:2401.14321}.

\bibitem[{Du et~al.(2024{\natexlab{b}})Du, Chen, Zhang, Hu et~al.}]{cosyvoice}
Zhihao Du, Qian Chen, Shiliang Zhang, Kai Hu, et~al. 2024{\natexlab{b}}.
\newblock Cosyvoice: A scalable multilingual zero-shot text-to-speech synthesizer based on supervised semantic tokens.
\newblock \emph{arXiv preprint arXiv:2407.05407}.

\bibitem[{Eskimez et~al.(2024)Eskimez, Wang, Thakker, Li et~al.}]{e2tts}
Sefik~Emre Eskimez, Xiaofei Wang, Manthan Thakker, Canrun Li, et~al. 2024.
\newblock {E2 TTS}: Embarrassingly easy fully non-autoregressive zero-shot {TTS}.
\newblock \emph{arXiv preprint arXiv:2406.18009}.

\bibitem[{Esser et~al.(2024)Esser, Kulal, Blattmann, Entezari et~al.}]{sd3}
Patrick Esser, Sumith Kulal, Andreas Blattmann, Rahim Entezari, et~al. 2024.
\newblock Scaling rectified flow transformers for high-resolution image synthesis.
\newblock In \emph{Proc. ICML}.

\bibitem[{Fei et~al.(2024)Fei, Fan, Yu, and Huang}]{fluxmusic}
Zhengcong Fei, Mingyuan Fan, Changqian Yu, and Junshi Huang. 2024.
\newblock Flux that plays music.
\newblock \emph{arXiv preprint arXiv:2409.00587}.

\bibitem[{Gao et~al.(2023{\natexlab{a}})Gao, Morioka, Zhang, and Chen}]{e3tts}
Yuan Gao, Nobuyuki Morioka, Yu~Zhang, and Nanxin Chen. 2023{\natexlab{a}}.
\newblock {E3 TTS}: Easy end-to-end diffusion-based text to speech.
\newblock In \emph{Proc. ASRU}, pages 1--8. IEEE.

\bibitem[{Gao et~al.(2023{\natexlab{b}})Gao, Li, Wang, Luo, Shi, Chen, Li, Zuo, Du, Xiao et~al.}]{funasr}
Zhifu Gao, Zerui Li, Jiaming Wang, Haoneng Luo, Xian Shi, Mengzhe Chen, Yabin Li, Lingyun Zuo, Zhihao Du, Zhangyu Xiao, et~al. 2023{\natexlab{b}}.
\newblock {FunASR}: A fundamental end-to-end speech recognition toolkit.
\newblock \emph{arXiv preprint arXiv:2305.11013}.

\bibitem[{Guo et~al.(2024{\natexlab{a}})Guo, Liu, Shen, Wu, Xie, Xie, and Xu}]{fireredtts}
Hao-Han Guo, Kun Liu, Fei-Yu Shen, Yi-Chen Wu, Feng-Long Xie, Kun Xie, and Kai-Tuo Xu. 2024{\natexlab{a}}.
\newblock {FireRedTTS}: A foundation text-to-speech framework for industry-level generative speech applications.
\newblock \emph{arXiv preprint arXiv:2409.03283}.

\bibitem[{Guo et~al.(2021)Guo, Wen, Jiang, Luo et~al.}]{didispeech}
Tingwei Guo, Cheng Wen, Dongwei Jiang, Ne~Luo, et~al. 2021.
\newblock Didispeech: A large scale {Mandarin} speech corpus.
\newblock In \emph{Proc. ICASSP}, pages 6968--6972. IEEE.

\bibitem[{Guo et~al.(2024{\natexlab{b}})Guo, Du, Ma, Chen, and Yu}]{voiceflow}
Yiwei Guo, Chenpeng Du, Ziyang Ma, Xie Chen, and Kai Yu. 2024{\natexlab{b}}.
\newblock Voiceflow: Efficient text-to-speech with rectified flow matching.
\newblock In \emph{Proc. ICASSP}, pages 11121--11125. IEEE.

\bibitem[{Han et~al.(2024)Han, Zhou, Liu, Chen, Meng, Qian, Liu, Zhao, Li, and Wei}]{valler}
Bing Han, Long Zhou, Shujie Liu, Sanyuan Chen, Lingwei Meng, Yanming Qian, Yanqing Liu, Sheng Zhao, Jinyu Li, and Furu Wei. 2024.
\newblock {VALL-E R}: Robust and efficient zero-shot text-to-speech synthesis via monotonic alignment.
\newblock \emph{arXiv preprint arXiv:2406.07855}.

\bibitem[{He et~al.(2024)He, Shang, Wang, Li et~al.}]{emilia}
Haorui He, Zengqiang Shang, Chaoren Wang, Xuyuan Li, et~al. 2024.
\newblock Emilia: An extensive, multilingual, and diverse speech dataset for large-scale speech generation.
\newblock \emph{arXiv preprint arXiv:2407.05361}.

\bibitem[{Ho et~al.(2020)Ho, Jain, and Abbeel}]{ddpm}
Jonathan Ho, Ajay Jain, and Pieter Abbeel. 2020.
\newblock Denoising diffusion probabilistic models.
\newblock \emph{Advances in neural information processing systems}, 33:6840--6851.

\bibitem[{Ho and Salimans(2022)}]{cfg}
Jonathan Ho and Tim Salimans. 2022.
\newblock Classifier-free diffusion guidance.
\newblock \emph{arXiv preprint arXiv:2207.12598}.

\bibitem[{Hsu et~al.(2021)Hsu, Bolte, Tsai, Lakhotia, Salakhutdinov, and Mohamed}]{hubert}
Wei-Ning Hsu, Benjamin Bolte, Yao-Hung~Hubert Tsai, Kushal Lakhotia, Ruslan Salakhutdinov, and Abdelrahman Mohamed. 2021.
\newblock Hubert: Self-supervised speech representation learning by masked prediction of hidden units.
\newblock \emph{IEEE/ACM Transactions on Audio, Speech, and Language Processing}, 29:3451--3460.

\bibitem[{Ito and Johnson(2017)}]{ljspeech}
Keith Ito and Linda Johnson. 2017.
\newblock \href {https://keithito.com/LJ-Speech-Dataset/} {The {LJ} speech dataset}.

\bibitem[{Ju et~al.(2024)Ju, Wang, Shen, Tan, Xin, Yang, Liu, Leng, Song, Tang et~al.}]{ns3}
Zeqian Ju, Yuancheng Wang, Kai Shen, Xu~Tan, Detai Xin, Dongchao Yang, Yanqing Liu, Yichong Leng, Kaitao Song, Siliang Tang, et~al. 2024.
\newblock Naturalspeech 3: Zero-shot speech synthesis with factorized codec and diffusion models.
\newblock \emph{arXiv preprint arXiv:2403.03100}.

\bibitem[{Kahn et~al.(2020)Kahn, Riviere, Zheng, Kharitonov et~al.}]{librilight}
Jacob Kahn, Morgane Riviere, Weiyi Zheng, Evgeny Kharitonov, et~al. 2020.
\newblock Libri-light: A benchmark for {ASR} with limited or no supervision.
\newblock In \emph{Proc. ICASSP}, pages 7669--7673. IEEE.

\bibitem[{Kang et~al.(2024)Kang, Yang, Yao, Kuang, Yang, Guo, Lin, and Povey}]{libriheavy}
Wei Kang, Xiaoyu Yang, Zengwei Yao, Fangjun Kuang, Yifan Yang, Liyong Guo, Long Lin, and Daniel Povey. 2024.
\newblock Libriheavy: a 50,000 hours asr corpus with punctuation casing and context.
\newblock In \emph{Proc. ICASSP}, pages 10991--10995. IEEE.

\bibitem[{Karras et~al.(2024)Karras, Aittala, Lehtinen, Hellsten, Aila, and Laine}]{ema}
Tero Karras, Miika Aittala, Jaakko Lehtinen, Janne Hellsten, Timo Aila, and Samuli Laine. 2024.
\newblock Analyzing and improving the training dynamics of diffusion models.
\newblock In \emph{Proceedings of the IEEE/CVF Conference on Computer Vision and Pattern Recognition}, pages 24174--24184.

\bibitem[{Kim et~al.(2020)Kim, Kim, Kong, and Yoon}]{glowtts}
Jaehyeon Kim, Sungwon Kim, Jungil Kong, and Sungroh Yoon. 2020.
\newblock {Glow-TTS}: A generative flow for text-to-speech via monotonic alignment search.
\newblock \emph{Advances in Neural Information Processing Systems}, 33:8067--8077.

\bibitem[{Kim et~al.(2021)Kim, Kong, and Son}]{vits}
Jaehyeon Kim, Jungil Kong, and Juhee Son. 2021.
\newblock Conditional variational autoencoder with adversarial learning for end-to-end text-to-speech.
\newblock In \emph{International Conference on Machine Learning}, pages 5530--5540. PMLR.

\bibitem[{Kingma and Gao(2024)}]{snrtheory}
Diederik Kingma and Ruiqi Gao. 2024.
\newblock Understanding diffusion objectives as the {ELBO} with simple data augmentation.
\newblock \emph{Advances in Neural Information Processing Systems}, 36.

\bibitem[{Le et~al.(2024)Le, Vyas, Shi, Karrer et~al.}]{voicebox}
Matthew Le, Apoorv Vyas, Bowen Shi, Brian Karrer, et~al. 2024.
\newblock Voicebox: Text-guided multilingual universal speech generation at scale.
\newblock \emph{Advances in neural information processing systems}, 36.

\bibitem[{Lee et~al.(2024)Lee, Kim, Kim, and Cho}]{dittotts}
Keon Lee, Dong~Won Kim, Jaehyeon Kim, and Jaewoong Cho. 2024.
\newblock {DiTTo-TTS}: Efficient and scalable zero-shot text-to-speech with diffusion transformer.
\newblock \emph{arXiv preprint arXiv:2406.11427}.

\bibitem[{Lee et~al.(2022)Lee, Ping, Ginsburg, Catanzaro, and Yoon}]{bigvgan}
Sang-gil Lee, Wei Ping, Boris Ginsburg, Bryan Catanzaro, and Sungroh Yoon. 2022.
\newblock Bigvgan: A universal neural vocoder with large-scale training.
\newblock \emph{arXiv preprint arXiv:2206.04658}.

\bibitem[{Li et~al.(2019)Li, Liu, Liu, Zhao, and Liu}]{transtts}
Naihan Li, Shujie Liu, Yanqing Liu, Sheng Zhao, and Ming Liu. 2019.
\newblock Neural speech synthesis with transformer network.
\newblock In \emph{Proceedings of the AAAI conference on artificial intelligence}, volume~33, pages 6706--6713.

\bibitem[{Li et~al.(2024)Li, Tian, Li, Deng, and He}]{ar-wovq}
Tianhong Li, Yonglong Tian, He~Li, Mingyang Deng, and Kaiming He. 2024.
\newblock Autoregressive image generation without vector quantization.
\newblock \emph{arXiv preprint arXiv:2406.11838}.

\bibitem[{Lipman et~al.(2022)Lipman, Chen, Ben-Hamu, Nickel, and Le}]{cfm-ot}
Yaron Lipman, Ricky~TQ Chen, Heli Ben-Hamu, Maximilian Nickel, and Matt Le. 2022.
\newblock Flow matching for generative modeling.
\newblock \emph{arXiv preprint arXiv:2210.02747}.

\bibitem[{Liu et~al.(2024{\natexlab{a}})Liu, Wang, Inoue, Bai, and Li}]{ardittts}
Zhijun Liu, Shuai Wang, Sho Inoue, Qibing Bai, and Haizhou Li. 2024{\natexlab{a}}.
\newblock Autoregressive diffusion transformer for text-to-speech synthesis.
\newblock \emph{arXiv preprint arXiv:2406.05551}.

\bibitem[{Liu et~al.(2024{\natexlab{b}})Liu, Wang, Zhu, Bi, and Li}]{e1tts}
Zhijun Liu, Shuai Wang, Pengcheng Zhu, Mengxiao Bi, and Haizhou Li. 2024{\natexlab{b}}.
\newblock {E1 TTS}: Simple and fast non-autoregressive {TTS}.
\newblock \emph{arXiv preprint arXiv:2409.09351}.

\bibitem[{Liu et~al.(2022)Liu, Mao, Wu, Feichtenhofer, Darrell, and Xie}]{convnext}
Zhuang Liu, Hanzi Mao, Chao-Yuan Wu, Christoph Feichtenhofer, Trevor Darrell, and Saining Xie. 2022.
\newblock A convnet for the 2020s.
\newblock In \emph{Proceedings of the IEEE/CVF conference on computer vision and pattern recognition}, pages 11976--11986.

\bibitem[{Loshchilov(2017)}]{adamw}
I~Loshchilov. 2017.
\newblock Decoupled weight decay regularization.
\newblock \emph{arXiv preprint arXiv:1711.05101}.

\bibitem[{Ma et~al.(2024)Ma, Guo, Song, Jiang, Wang, Xue, Xu, Zhao, Zhang, and Xie}]{wenet4tts}
Linhan Ma, Dake Guo, Kun Song, Yuepeng Jiang, Shuai Wang, Liumeng Xue, Weiming Xu, Huan Zhao, Binbin Zhang, and Lei Xie. 2024.
\newblock {WenetSpeech4TTS}: A 12,800-hour {Mandarin TTS} corpus for large speech generation model benchmark.
\newblock \emph{arXiv preprint arXiv:2406.05763}.

\bibitem[{Mehta et~al.(2024)Mehta, Tu, Beskow, Sz{\'e}kely, and Henter}]{matchatts}
Shivam Mehta, Ruibo Tu, Jonas Beskow, {\'E}va Sz{\'e}kely, and Gustav~Eje Henter. 2024.
\newblock {Matcha-TTS}: A fast {TTS} architecture with conditional flow matching.
\newblock In \emph{Proc. ICASSP}, pages 11341--11345. IEEE.

\bibitem[{Meister et~al.(2023)Meister, Novikov, Karpov, Bakhturina, Lavrukhin, and Ginsburg}]{librispeechpc}
Aleksandr Meister, Matvei Novikov, Nikolay Karpov, Evelina Bakhturina, Vitaly Lavrukhin, and Boris Ginsburg. 2023.
\newblock {LibriSpeech-PC}: Benchmark for evaluation of punctuation and capitalization capabilities of end-to-end {ASR} models.
\newblock In \emph{Proc. ASRU}, pages 1--7. IEEE.

\bibitem[{Meng et~al.(2024)Meng, Zhou, Liu, Chen et~al.}]{melle}
Lingwei Meng, Long Zhou, Shujie Liu, Sanyuan Chen, et~al. 2024.
\newblock Autoregressive speech synthesis without vector quantization.
\newblock \emph{arXiv preprint arXiv:2407.08551}.

\bibitem[{Niu et~al.(2024)Niu, Chen, Zhou, Ma, Chen, and Liu}]{ndvq}
Zhikang Niu, Sanyuan Chen, Long Zhou, Ziyang Ma, Xie Chen, and Shujie Liu. 2024.
\newblock {NDVQ}: Robust neural audio codec with normal distribution-based vector quantization.
\newblock \emph{arXiv preprint arXiv:2409.12717}.

\bibitem[{Okamoto et~al.(2024)Okamoto, Ohtani, Toda, and Kawai}]{convnexttts}
Takuma Okamoto, Yamato Ohtani, Tomoki Toda, and Hisashi Kawai. 2024.
\newblock {Convnext-TTS} and {Convnext-VC}: Convnext-based fast end-to-end sequence-to-sequence text-to-speech and voice conversion.
\newblock In \emph{Proc. ICASSP}, pages 12456--12460. IEEE.

\bibitem[{Peebles and Xie(2023)}]{dit}
William Peebles and Saining Xie. 2023.
\newblock Scalable diffusion models with transformers.
\newblock In \emph{Proceedings of the IEEE/CVF International Conference on Computer Vision}, pages 4195--4205.

\bibitem[{Peng et~al.(2024)Peng, Huang, Li, Mohamed, and Harwath}]{voicecraft}
Puyuan Peng, Po-Yao Huang, Daniel Li, Abdelrahman Mohamed, and David Harwath. 2024.
\newblock Voicecraft: Zero-shot speech editing and text-to-speech in the wild.
\newblock \emph{arXiv preprint arXiv:2403.16973}.

\bibitem[{Popov et~al.(2021)Popov, Vovk, Gogoryan, Sadekova, and Kudinov}]{gradtts}
Vadim Popov, Ivan Vovk, Vladimir Gogoryan, Tasnima Sadekova, and Mikhail Kudinov. 2021.
\newblock Grad-tts: A diffusion probabilistic model for text-to-speech.
\newblock In \emph{International Conference on Machine Learning}, pages 8599--8608. PMLR.

\bibitem[{Press et~al.(2021)Press, Smith, and Lewis}]{alibi}
Ofir Press, Noah~A Smith, and Mike Lewis. 2021.
\newblock Train short, test long: Attention with linear biases enables input length extrapolation.
\newblock \emph{arXiv preprint arXiv:2108.12409}.

\bibitem[{Radford et~al.(2023)Radford, Kim, Xu, Brockman, McLeavey, and Sutskever}]{whisper}
Alec Radford, Jong~Wook Kim, Tao Xu, Greg Brockman, Christine McLeavey, and Ilya Sutskever. 2023.
\newblock Robust speech recognition via large-scale weak supervision.
\newblock In \emph{International conference on machine learning}, pages 28492--28518. PMLR.

\bibitem[{Ren et~al.(2020)Ren, Hu, Tan, Qin, Zhao, Zhao, and Liu}]{fastspeech2}
Yi~Ren, Chenxu Hu, Xu~Tan, Tao Qin, Sheng Zhao, Zhou Zhao, and Tie-Yan Liu. 2020.
\newblock Fastspeech 2: Fast and high-quality end-to-end text to speech.
\newblock \emph{arXiv preprint arXiv:2006.04558}.

\bibitem[{Ronneberger et~al.(2015)Ronneberger, Fischer, and Brox}]{unet}
Olaf Ronneberger, Philipp Fischer, and Thomas Brox. 2015.
\newblock U-net: Convolutional networks for biomedical image segmentation.
\newblock In \emph{Proc. MICCAI}, pages 234--241. Springer.

\bibitem[{Saeki et~al.(2022)Saeki, Xin, Nakata, Koriyama, Takamichi, and Saruwatari}]{utmos}
Takaaki Saeki, Detai Xin, Wataru Nakata, Tomoki Koriyama, Shinnosuke Takamichi, and Hiroshi Saruwatari. 2022.
\newblock Utmos: Utokyo-sarulab system for voicemos challenge 2022.
\newblock \emph{arXiv preprint arXiv:2204.02152}.

\bibitem[{Salimans and Ho(2022)}]{vpredict}
Tim Salimans and Jonathan Ho. 2022.
\newblock Progressive distillation for fast sampling of diffusion models.
\newblock \emph{arXiv preprint arXiv:2202.00512}.

\bibitem[{Shen et~al.(2018)Shen, Pang, Weiss, Schuster et~al.}]{tacotron2}
Jonathan Shen, Ruoming Pang, Ron~J Weiss, Mike Schuster, et~al. 2018.
\newblock Natural {TTS} synthesis by conditioning wavenet on mel spectrogram predictions.
\newblock In \emph{Proc. ICASSP}, pages 4779--4783. IEEE.

\bibitem[{Shen et~al.(2023)Shen, Ju, Tan, Liu et~al.}]{ns2}
Kai Shen, Zeqian Ju, Xu~Tan, Yanqing Liu, et~al. 2023.
\newblock Naturalspeech 2: Latent diffusion models are natural and zero-shot speech and singing synthesizers.
\newblock \emph{arXiv preprint arXiv:2304.09116}.

\bibitem[{Siuzdak(2023)}]{vocos}
Hubert Siuzdak. 2023.
\newblock Vocos: Closing the gap between time-domain and fourier-based neural vocoders for high-quality audio synthesis.
\newblock \emph{arXiv preprint arXiv:2306.00814}.

\bibitem[{Song et~al.(2024)Song, Chen, Wang, Ma, and Chen}]{ellav}
Yakun Song, Zhuo Chen, Xiaofei Wang, Ziyang Ma, and Xie Chen. 2024.
\newblock {ELLA-V}: Stable neural codec language modeling with alignment-guided sequence reordering.
\newblock \emph{arXiv preprint arXiv:2401.07333}.

\bibitem[{Song et~al.(2020)Song, Sohl-Dickstein, Kingma, Kumar, Ermon, and Poole}]{score}
Yang Song, Jascha Sohl-Dickstein, Diederik~P Kingma, Abhishek Kumar, Stefano Ermon, and Ben Poole. 2020.
\newblock Score-based generative modeling through stochastic differential equations.
\newblock \emph{arXiv preprint arXiv:2011.13456}.

\bibitem[{Su et~al.(2024)Su, Ahmed, Lu, Pan, Bo, and Liu}]{rope}
Jianlin Su, Murtadha Ahmed, Yu~Lu, Shengfeng Pan, Wen Bo, and Yunfeng Liu. 2024.
\newblock Roformer: Enhanced transformer with rotary position embedding.
\newblock \emph{Neurocomputing}, 568:127063.

\bibitem[{Tan et~al.(2024)Tan, Chen, Liu, Cong et~al.}]{ns1}
Xu~Tan, Jiawei Chen, Haohe Liu, Jian Cong, et~al. 2024.
\newblock Naturalspeech: End-to-end text-to-speech synthesis with human-level quality.
\newblock \emph{IEEE Transactions on Pattern Analysis and Machine Intelligence}.

\bibitem[{Wang et~al.(2023{\natexlab{a}})Wang, Chen, Wu, Zhang et~al.}]{valle}
Chengyi Wang, Sanyuan Chen, Yu~Wu, Ziqiang Zhang, et~al. 2023{\natexlab{a}}.
\newblock Neural codec language models are zero-shot text to speech synthesizers.
\newblock \emph{arXiv preprint arXiv:2301.02111}.

\bibitem[{Wang et~al.(2023{\natexlab{b}})Wang, Zhou, Zhang, Wu, Liu, Gaur, Chen, Li, and Wei}]{viola}
Tianrui Wang, Long Zhou, Ziqiang Zhang, Yu~Wu, Shujie Liu, Yashesh Gaur, Zhuo Chen, Jinyu Li, and Furu Wei. 2023{\natexlab{b}}.
\newblock {VioLA}: Unified codec language models for speech recognition, synthesis, and translation.
\newblock \emph{arXiv preprint arXiv:2305.16107}.

\bibitem[{Wang et~al.(2024)Wang, Zhan, Liu, Zeng, Guo, Zheng, Zhang, Zhang, and Wu}]{maskgct}
Yuancheng Wang, Haoyue Zhan, Liwei Liu, Ruihong Zeng, Haotian Guo, Jiachen Zheng, Qiang Zhang, Shunsi Zhang, and Zhizheng Wu. 2024.
\newblock {MaskGCT}: Zero-shot text-to-speech with masked generative codec transformer.
\newblock \emph{arXiv preprint arXiv:2409.00750}.

\bibitem[{Woo et~al.(2023)Woo, Debnath, Hu, Chen, Liu, Kweon, and Xie}]{convnextv2}
Sanghyun Woo, Shoubhik Debnath, Ronghang Hu, Xinlei Chen, Zhuang Liu, In~So Kweon, and Saining Xie. 2023.
\newblock Convnext v2: Co-designing and scaling convnets with masked autoencoders.
\newblock In \emph{Proceedings of the IEEE/CVF Conference on Computer Vision and Pattern Recognition}, pages 16133--16142.

\bibitem[{Wu et~al.(2023)Wu, Gebru, Markovi{\'c}, and Richard}]{audiodec}
Yi-Chiao Wu, Israel~D Gebru, Dejan Markovi{\'c}, and Alexander Richard. 2023.
\newblock Audiodec: An open-source streaming high-fidelity neural audio codec.
\newblock In \emph{Proc. ICASSP}, pages 1--5. IEEE.

\bibitem[{Xin et~al.(2024)Xin, Tan, Shen, Ju et~al.}]{ralle}
Detai Xin, Xu~Tan, Kai Shen, Zeqian Ju, et~al. 2024.
\newblock Rall-e: Robust codec language modeling with chain-of-thought prompting for text-to-speech synthesis.
\newblock \emph{arXiv preprint arXiv:2404.03204}.

\bibitem[{Yang et~al.(2023)Yang, Liu, Huang, Tian, Weng, and Zou}]{hificodec}
Dongchao Yang, Songxiang Liu, Rongjie Huang, Jinchuan Tian, Chao Weng, and Yuexian Zou. 2023.
\newblock Hifi-codec: Group-residual vector quantization for high fidelity audio codec.
\newblock \emph{arXiv preprint arXiv:2305.02765}.

\bibitem[{Zeghidour et~al.(2021)Zeghidour, Luebs, Omran, Skoglund, and Tagliasacchi}]{soundstream}
Neil Zeghidour, Alejandro Luebs, Ahmed Omran, Jan Skoglund, and Marco Tagliasacchi. 2021.
\newblock Soundstream: An end-to-end neural audio codec.
\newblock \emph{IEEE/ACM Transactions on Audio, Speech, and Language Processing}, 30:495--507.

\bibitem[{Zen et~al.(2019)Zen, Dang, Clark, Zhang, Weiss, Jia, Chen, and Wu}]{libritts}
Heiga Zen, Viet Dang, Rob Clark, Yu~Zhang, Ron~J Weiss, Ye~Jia, Zhifeng Chen, and Yonghui Wu. 2019.
\newblock Libritts: A corpus derived from librispeech for text-to-speech.
\newblock \emph{arXiv preprint arXiv:1904.02882}.

\bibitem[{Zhang et~al.(2023{\natexlab{a}})Zhang, Zhang, Li, Zhou, and Qiu}]{speechtokenizer}
Xin Zhang, Dong Zhang, Shimin Li, Yaqian Zhou, and Xipeng Qiu. 2023{\natexlab{a}}.
\newblock Speechtokenizer: Unified speech tokenizer for speech large language models.
\newblock \emph{arXiv preprint arXiv:2308.16692}.

\bibitem[{Zhang et~al.(2023{\natexlab{b}})Zhang, Zhou, Wang, Chen et~al.}]{vallex}
Ziqiang Zhang, Long Zhou, Chengyi Wang, Sanyuan Chen, et~al. 2023{\natexlab{b}}.
\newblock Speak foreign languages with your own voice: Cross-lingual neural codec language modeling.
\newblock \emph{arXiv preprint arXiv:2303.03926}.

\end{thebibliography}

\clearpage

\appendix

\section{Baseline Details}
\label{appx:baselines}
\textbf{VALL-E 2}~\citep{valle2}\quad A large-scale TTS model shares the same architecture as VALL-E~\citep{valle} but employs a repetition-aware sampling strategy that promotes more deliberate sampling choices, trained on Libriheavy~\citep{libriheavy} 50K hours English dataset. We compared with results reported in~\citet{melle}.
\\[-0.6em]

\noindent
\textbf{MELLE}~\citep{melle}\quad An autoregressive large-scale model leverages continuous-valued tokens with variational inference for text-to-speech synthesis. Its variants allow to prediction of multiple mel-spectrogram frames at each time step, noted by MELLE-\textit{Rx} with \textit{x} denotes reduction factor. The model is trained on Libriheavy~\citep{libriheavy} 50K hours English dataset. We compared with results reported in~\citet{melle}.
\\[-0.6em]

\noindent
\textbf{Voicebox}~\citep{voicebox}\quad A non-autoregressive large-scale model based on flow matching trained with infilling task. We compared with the 330M parameters trained on 60K hours dataset English-only model's results reported in~\citet{voicebox} and ~\citet{ns3}.
\\[-0.6em]

\noindent
\textbf{NaturalSpeech 3}~\citep{ns3}\quad A non-autoregressive large-scale TTS system leverages a factorized neural codec to decouple speech representations and a factorized diffusion model to generate speech based on disentangled attributes. The 500M base model is trained on Librilight~\citep{librilight} a 60K hours English dataset. We compared with scores reported in~\citet{ns3}.
\\[-0.6em]

\noindent
\textbf{DiTTo-TTS}~\citep{dittotts}\quad A large-scale non-autoregressive TTS model uses a cross-attention Diffusion Transformer and leverages a pretrained language model to enhance the alignment. We compare with DiTTo-en-XL, a 740M model trained on 55K hours English-only dataset, using scores reported in~\citet{dittotts}.
\\[-0.6em]

\noindent
\textbf{FireRedTTS}~\citep{fireredtts}\quad A foundation TTS framework for industry-level generative speech applications. The autoregressive text-to-semantic token model has 400M parameters and the token-to-waveform generation model has about half the parameters. The system is trained with 248K hours of labeled speech data. We use the official code and pre-trained checkpoint to evaluate\footnote{\url{https://github.com/FireRedTeam/FireRedTTS}}.
\\[-0.6em]

\noindent
\textbf{MaskGCT}~\citep{maskgct}\quad A large-scale non-autoregressive TTS model without precise alignment information between text and speech following the mask-and-predict learning paradigm. The model is multi-stage, with a 695M text-to-semantic model (T2S) and then a 353M semantic-to-acoustic (S2A) model. The model is trained on Emilia~\citep{emilia} dataset with around 100K Chinese and English in-the-wild speech data. We compare with results reported in~\citet{maskgct}.
\\[-0.6em]

\noindent
\textbf{Seed-TTS}~\citep{seedtts}\quad A family of high-quality versatile speech generation models trained on unknown tremendously large data that is of orders of magnitudes larger than the previously largest TTS systems~\citep{seedtts}. Seed-TTS$_{DiT}$ is a large-scale fully non-autoregressive model. We compare with results reported in~\citet{seedtts}.
\\[-0.6em]

\noindent
\textbf{E2 TTS}~\citep{e2tts}\quad A fully non-autoregressive TTS system proposes to model without the phoneme-level alignment in Voicebox, originally trained on Libriheavy~\citep{libriheavy} 50K English dataset. We compare with our reproduced 333M multilingual E2 TTS trained on Emilia~\citep{emilia} dataset with around 100K Chinese and English in-the-wild speech data.
\\[-0.6em]

\noindent
\textbf{CosyVoice}~\citep{cosyvoice}\quad A two-stage large-scale TTS system, first autoregressive text-to-token, then a flow matching diffusion model. The model is of around 300M parameters, trained on 170K hours of multilingual speech data. We obtain the evaluation result with the official code and pre-trained checkpoint\footnote{\url{https://huggingface.co/model-scope/CosyVoice-300M}}.

\section{Experimental Result Supplements}

The UTMOS \citep{utmos} scores reported in this section are evaluated with an open-source MOS prediction model\footnote{\url{https://github.com/tarepan/SpeechMOS}}. The UTMOS is an objective metric measuring naturalness.

\subsection{Small Model Configuration}
\label{appx:smallmodels}

The detailed configuration of small models is shown in Tab.\ref{tab:smallmodels}. In the Transformer column, the numbers denote the Model Dimension, the Number of Layers, the Number of Heads, and the multiples of Hidden Size. In the ConvNeXt column, the numbers denote the Model Dimension, the Number of Layers, and the multiples of Hidden Size. GFLOPs are evaluated using the \texttt{thop} Python package.

As mentioned in Sec.\ref{sec:f5tts}, F5-TTS leverages an adaLN DiT backbone with ConvNeXt V2 blocks, while E2 TTS is a flat U-Net equipped Transformer. F5-TTS\textit{$+$LongSkip} adds an additional long skip structure connecting the first to the last layer \citep{dittotts} in the Transformer. For the Multi-Model Diffusion Transformer (MMDiT) \citep{sd3}, a double stream transformer, the setting denotes one stream configuration.

\begin{table}[h]
\begin{center}
\resizebox{1\linewidth}{!}{
\begin{tabular}{lccrc}
\toprule
\multicolumn{5}{l}{\bf Model \hspace{5.2em} \bf Transformer \bf ConvNeXt \bf \#Param. \bf GFLOPs} \\
\midrule
\multicolumn{2}{l}{F5-TTS                      \hspace{5.7em} 768,18,12,2}  &512,4,2  &158M  &173 \\
\multicolumn{2}{l}{F5-TTS\textit{$-$Conv2Text} \hspace{0.7em} 768,18,12,2}  &-        &153M  &164 \\
\multicolumn{2}{l}{F5-TTS\textit{$+$Conv2Audio}\hspace{0.2em} 768,16,12,2}  &512,4,2  &163M  &181 \\
\multicolumn{2}{l}{F5-TTS\textit{$+$LongSkip}  \hspace{1.1em} 768,18,12,2}  &512,4,2  &159M  &175 \\
\multicolumn{2}{l}{E2 TTS                      \hspace{5.7em} 768,20,12,4}  &-        &157M  &293 \\
\multicolumn{2}{l}{E2 TTS\textit{$+$Conv2Text} \hspace{0.7em} 768,20,12,4}  &512,4,2  &161M  &301 \\
\multicolumn{2}{l}{MMDiT                       \hspace{5.4em} 512,16,16,2}  &-        &151M  &104 \\
\bottomrule
\end{tabular}
}
\end{center}
\caption{Details of small model configurations.}
\label{tab:smallmodels}
\end{table}

\subsection{Ablation study on Input Condition}
\label{appx:ablationinput}

The ablation study on different input conditions is conducted with three settings: 
\begin{itemize}
    \item Common input with text and audio prompts.
    \item Providing ground truth duration information rather than an estimate.
    \item Retaining only text input, dropping audio prompt (using blank).
\end{itemize}
In Tab.\ref{tab:ablationinput}, all evaluations take the 155M small models' checkpoints trained on WenetSpeech4TTS Premium at 800K updates. Analysis see Sec.\ref{sec:modelarchitecture}.

\begin{table}[h]
\begin{center}
\resizebox{1\linewidth}{!}{
\begin{tabular}{lrr|rr|rr}
\toprule
\multicolumn{7}{l}{\bf Model \hspace{4.0em} \bf Common Input \hspace{0.45em} \bf GT Duration \hspace{0.5em} \bf Text-Only { }{ }} \\
\multicolumn{7}{r}{\bf WER↓ \bf SIM↑ \ \bf WER↓ \bf SIM↑ \ \bf WER↓ SIM↑} \\
\midrule
\multicolumn{2}{l}{F5-TTS                      \hspace{5.7em} \textbf{4.17}}     &\underline{0.54}  &\textbf{3.87}     &\underline{0.54}  &\underline{3.22}  &0.21 \\
\multicolumn{2}{l}{F5-TTS\textit{$+$Conv2Audio}\hspace{0.2em} 5.78}              &\textbf{0.55}     &5.28              &\textbf{0.55}     &3.78              &0.21 \\
\multicolumn{2}{l}{F5-TTS\textit{$+$LongSkip}  \hspace{1.1em} \underline{5.17}}  &0.53              &\underline{5.03}  &0.53              &3.35              &0.21 \\
\multicolumn{2}{l}{E2 TTS                      \hspace{5.7em} 9.63}              &0.53              &9.48              &0.53              &3.48              &0.21 \\
\multicolumn{2}{l}{E2 TTS\textit{$+$Conv2Text} \hspace{0.2em} 18.10}             &0.49              &17.94             &0.49              &\textbf{3.06}     &0.21 \\
\bottomrule
\end{tabular}
}
\end{center}
\caption{Ablation study on different input conditions. The boldface indicates the best result, and the underline denotes the second best. All scores are the average of three random seed results.}
\label{tab:ablationinput}
\end{table}

\subsection{Sway Sampling Effectiveness on Base Models}
\label{appx:swaysampling}

From Tab.\ref{tab:swaysampling}, it is clear that our Sway Sampling strategy for test-time flow steps consistently improves the zero-shot generation performance in aspects of faithfulness to prompt text (WER), speaker similarity (SIM), and naturalness (UTMOS). The gain of applying Sway Sampling to E2 TTS \citep{e2tts} proves that our Sway Sampling strategy is universally applicable to existing flow matching based TTS models.

\begin{table}[h]
\begin{center}
\resizebox{1\linewidth}{!}{
\begin{tabular}{lcccc}
\toprule
\multicolumn{2}{l}{\bf Model \hspace{5.4em} \bf WER(\%)↓} &\bf SIM-o↑ &\bf UTMOS↑ &\bf RTF↓ \\
\midrule
\multicolumn{5}{c}{\textbf{LibriSpeech-PC \textit{test-clean}}} \\
\midrule
Ground Truth                                  &2.23  &0.69  &4.09  &-  \\
Vocoder Resynthesized                         &2.32  &0.66  &3.64  &-  \\
\cdashline{1-5}\noalign{\vskip\belowrulesep}
E2 TTS (16 NFE \textit{w/} SS)                &2.86  &0.71  &3.66  &0.34\\
E2 TTS (32 NFE \textit{w/} SS)                &2.84  &0.72  &3.70  &0.68\\
E2 TTS (32 NFE \textit{w/o} SS)               &2.95  &0.69  &3.56  &0.68\\
\cdashline{1-5}\noalign{\vskip\belowrulesep}
F5-TTS (16 NFE \textit{w/} SS)                &2.43  &0.66  &3.87  &0.26\\
F5-TTS (32 NFE \textit{w/} SS)                &2.41  &0.66  &3.89  &0.53\\
F5-TTS (32 NFE \textit{w/o} SS)               &2.84  &0.62  &3.70  &0.53\\
\midrule
\multicolumn{5}{c}{\textbf{Seed-TTS \textit{test-en}}} \\
\midrule
Ground Truth                                  &2.06  &0.73  &3.53  &-  \\
Vocoder Resynthesized                         &2.09  &0.70  &3.33  &-  \\
\cdashline{1-5}\noalign{\vskip\belowrulesep}
E2 TTS (16 NFE \textit{w/} SS)                &1.99  &0.72  &3.55  &0.34\\
E2 TTS (32 NFE \textit{w/} SS)                &1.98  &0.73  &3.57  &0.68\\
E2 TTS (32 NFE \textit{w/o} SS)               &2.19  &0.71  &3.33  &0.68\\
\cdashline{1-5}\noalign{\vskip\belowrulesep}
F5-TTS (16 NFE \textit{w/} SS)                &1.88  &0.66  &3.70  &0.26\\
F5-TTS (32 NFE \textit{w/} SS)                &1.87  &0.66  &3.72  &0.53\\
F5-TTS (32 NFE \textit{w/o} SS)               &1.93  &0.63  &3.51  &0.53\\
\midrule
\multicolumn{5}{c}{\textbf{Seed-TTS \textit{test-zh}}} \\
\midrule
Ground Truth                                  &1.26  &0.76  &2.78  &-  \\
Vocoder Resynthesized                         &1.27  &0.72  &2.61  &-  \\
\cdashline{1-5}\noalign{\vskip\belowrulesep}
E2 TTS (16 NFE \textit{w/} SS)                &1.80  &0.78  &2.84  &0.34\\
E2 TTS (32 NFE \textit{w/} SS)                &1.77  &0.78  &2.87  &0.68\\
E2 TTS (32 NFE \textit{w/o} SS)               &1.97  &0.73  &2.49  &0.68\\
\cdashline{1-5}\noalign{\vskip\belowrulesep}
F5-TTS (16 NFE \textit{w/} SS)                &1.61  &0.75  &2.87  &0.26\\
F5-TTS (32 NFE \textit{w/} SS)                &1.58  &0.75  &2.91  &0.53\\
F5-TTS (32 NFE \textit{w/o} SS)               &1.93  &0.69  &2.58  &0.53\\
\bottomrule
\end{tabular}
}
\end{center}
\caption{Base model evaluation results on LibriSpeech-PC \textit{test-clean}, Seed-TTS \textit{test-en} and \textit{test-zh}, with and without proposed test-time Sway Sampling (SS, with coefficient $s=-1$) strategy for flow steps. All generations leverage the midpoint ODE solver for ease of ablation.}
\label{tab:swaysampling}
\end{table}

\subsection{Comparison of ODE Solvers}
\label{appx:odesolver}

The comparison results of using the Euler (first-order), midpoint (second-order), or improved Heun (third-order, Heun-3) ODE solver during F5-TTS inference are shown in Tab.\ref{tab:odesolver}. The Euler is inherently faster and performs slightly better typically for larger NFE inference with Sway Sampling (otherwise the Euler solver results in degradation). 

\begin{table*}[ht]
\begin{center}
\resizebox{1\linewidth}{!}{
\begin{tabular}{lccc|ccc|ccc|c}
\toprule
{ } &\multicolumn{3}{c}{\bf LibriSpeech-PC \textit{test-clean}} &\multicolumn{3}{c}{\textbf{Seed-TTS \textit{test-en}}} &\multicolumn{3}{c}{\textbf{Seed-TTS \textit{test-zh}}} &{ }\\
\bf F5-TTS &\bf WER(\%)↓ &\bf SIM-o↑ &\bf UTMOS↑ &\bf WER(\%)↓ &\bf SIM-o↑ &\bf UTMOS↑ &\bf WER(\%)↓ &\bf SIM-o↑ &\bf UTMOS↑ &\bf { } RTF↓ { } \\
\midrule
Ground Truth    &2.23 &0.69 &4.09{ }  &{ }2.06 &0.73 &3.53  &{ }1.26 &0.76 &2.78  &-    \\
\cdashline{1-11}\noalign{\vskip\belowrulesep}
$s=-1$          &     &     &         &        &     &      &        &     &      &     \\
16 NFE Euler    &2.53 &0.66 &3.88{ }  &{ }1.89 &0.67 &3.76  &{ }1.74 &0.75 &2.96  &0.15 \\
16 NFE midpoint &2.43 &0.66 &3.87{ }  &{ }1.88 &0.66 &3.70  &{ }1.61 &0.75 &2.87  &0.26 \\
32 NFE Euler    &2.42 &0.66 &3.90{ }  &{ }1.83 &0.67 &3.76  &{ }1.56 &0.76 &2.95  &0.31 \\
32 NFE midpoint &2.41 &0.66 &3.89{ }  &{ }1.87 &0.66 &3.72  &{ }1.58 &0.75 &2.91  &0.53 \\
16 NFE Heun-3   &2.39 &0.65 &3.87{ }  &{ }1.80 &0.66 &3.70  &{ }1.55 &0.75 &2.88  &0.44 \\
\cdashline{1-11}\noalign{\vskip\belowrulesep}
$s=-0.8$        &     &     &         &        &     &      &        &     &      &     \\
16 NFE Euler    &2.82 &0.65 &3.73{ }  &{ }2.14 &0.65 &3.70  &{ }2.28 &0.72 &2.74  &0.15 \\
16 NFE midpoint &2.58 &0.65 &3.86{ }  &{ }1.86 &0.65 &3.68  &{ }1.70 &0.73 &2.83  &0.26 \\
32 NFE Euler    &2.50 &0.66 &3.89{ }  &{ }1.81 &0.67 &3.74  &{ }1.62 &0.75 &2.94  &0.31 \\
32 NFE midpoint &2.42 &0.66 &3.89{ }  &{ }1.84 &0.66 &3.70  &{ }1.62 &0.75 &2.91  &0.53 \\
16 NFE Heun-3   &2.40 &0.65 &3.85{ }  &{ }1.78 &0.66 &3.68  &{ }1.56 &0.74 &2.84  &0.44 \\
\bottomrule
\end{tabular}
}
\end{center}
\caption{Evaluation results of F5-TTS on LibriSpeech-PC \textit{test-clean}, Seed-TTS \textit{test-en} and Seed-TTS \textit{test-zh}, employing the Euler, midpoint or Heun-3 ODE solver, and with different Sway Sampling $s$ values.}
\label{tab:odesolver}
\end{table*}

\subsection{ELLA-V Hard Sentences Evaluation}
\label{appx:ellavhardtest}

ELLA-V \citep{ellav} proposed a challenging set containing 100 difficult textual patterns evaluating the robustness of the TTS model. Following previous works \citep{valle2,melle,e2tts}, we include generated samples in our demo page\footnote{\url{https://SWivid.github.io/F5-TTS/}}. We additionally compare our model with the objective evaluation results reported in E1 TTS \citep{e1tts}.

StyleTTS 2 is a TTS model leveraging style diffusion and adversarial training with large speech language models. CosyVoice is a two-stage large-scale TTS system, consisting of a text-to-token AR model and a token-to-speech flow matching model. Concurrent with our work, E1 TTS$_{DMD}$ is a diffusion-based NAR model with a distribution matching distillation technique to achieve one-step TTS generation.  
Since the prompts used by E1 TTS$_{DMD}$ are not released, we randomly sample 3-second-long speeches in our LibriSpeech-PC \textit{test-clean} set as audio prompts. 
The evaluation result is in Tab.\ref{tab:ellavhardtest}. We evaluate the reproduced E2 TTS and our F5-TTS with 32 NFE and Sway Sampling and report the averaged score of three random seed results. 

\begin{table}[ht]
\begin{center}
\resizebox{1\linewidth}{!}{
\begin{tabular}{lcccc}
\toprule
{\bf Model} &{\bf WER(\%)↓} &{\bf Sub.(\%)↓} &{\bf Del.(\%)↓} &{\bf Ins.(\%)↓} \\
\midrule
StyleTTS 2     &4.83*       &2.17*       &2.03*       &0.61* \\
CosyVoice      &8.30*       &3.47*       &2.74*       &1.93* \\
E1 TTS$_{DMD}$ &4.29*       &1.89*       &1.62*       &0.74* \\
E2 TTS         &8.58{ }{ }  &3.70{ }{ }  &4.82{ }{ }  &0.06{ }{ } \\
F5-TTS         &4.40{ }{ }  &1.81{ }{ }  &2.40{ }{ }  &0.18{ }{ } \\
\bottomrule
\end{tabular}
}
\end{center}
\caption{Results of zero-shot TTS WER on ELLA-V hard sentences. The asterisk * denotes the score reported in E1 TTS. Sub. for Substitution, Del. for Deletion, and Ins. for Insertion.}
\label{tab:ellavhardtest}
\end{table}

We note that a higher WER compared to the results on commonly used test sets is partially due to mispronunciation (\textit{yogis} to \textit{yojus}, \textit{cavorts} to \textit{caverts}, \textit{etc.}). The high Deletion rate indicates a word-skipping phenomenon when our model encounters a stack of repeating words. The low Insertion rate makes it clear that our model is free of endless repetition. We further emphasize that prompts from different speakers will spell very distinct utterances, where the ASR model transcribes correctly for one, and fails for another (\textit{e.g.} \textit{quokkas} to \textit{Cocos}).

\subsection{Comparison of Vocoders and between PC and non-PC}
\label{appx:compr_vocoder}

The inference results with pretrained BigVGAN \citep{bigvgan} and Vocos \citep{vocos} respectively as vocoder are shown in Tab.\ref{tab:compr_vocoder}, along with additional evaluation on a non-Capitalized version removing all Punctuations (non-PC) of the filtered LibriSpeech-PC \textit{test-clean} subset. The non-PC version equals an ordinary  LibriSpeech \textit{test-clean} subset, with which we provide more comprehensive comparisons with previous works. 

Moreover, we include WER scores measuring with a Hubert-large-based \citep{hubert} ASR model\footnote{\url{https://huggingface.co/facebook/hubert-large-ls960-ft}} in Tab.\ref{tab:compr_vocoder}, with which our reproduced multilingual E2 TTS with 32 NFE and Vocos as vocoder achieves a WER of 2.92 on LibriSpeech-PC \textit{test-clean} and 2.66 if Sway Sampling applied.

\begin{table*}[ht]
\begin{center}
\resizebox{1\linewidth}{!}{
\begin{tabular}{lcc|cc|cc|cc}
\toprule
\bf NFE steps &\multicolumn{2}{c}{\bf LibriSpeech-PC \textit{test-clean}} &\multicolumn{2}{c}{\bf LibriSpeech-non-PC \textit{test-clean}} &\multicolumn{2}{c}{\textbf{Seed-TTS \textit{test-en}}} &\multicolumn{2}{c}{\textbf{Seed-TTS \textit{test-zh}}}\\
\bf\& Vocoder &\bf WER(\%)↓ &\bf SIM-o ↑ &\bf WER(\%)↓ &\bf SIM-o ↑ &\bf WER(\%)↓ &\bf SIM-o ↑ &\bf WER(\%)↓ &\bf SIM-o ↑ \\
\midrule
Ground Truth      &{ } 2.23 { } (1.89) &0.69    &{ } 2.29 { } (1.86) &0.69    &2.06 &0.73    &1.26  &0.76  \\
\cdashline{1-9}\noalign{\vskip\belowrulesep}
16 NFE - Vocos    &{ } 2.53 { } (2.34) &0.66    &{ } 2.72 { } (2.53) &0.66    &1.89 &0.67    &1.74  &0.75 \\
16 NFE - BigVGAN  &{ } 2.21 { } (1.96) &0.67    &{ } 2.55 { } (2.34) &0.67    &1.65 &0.66    &1.64  &0.74 \\
32 NFE - Vocos    &{ } 2.42 { } (2.09) &0.66    &{ } 2.44 { } (2.16) &0.66    &1.83 &0.67    &1.56  &0.76 \\
32 NFE - BigVGAN  &{ } 2.11 { } (1.81) &0.67    &{ } 2.28 { } (2.03) &0.67    &1.62 &0.66    &1.53  &0.74 \\
\bottomrule
\end{tabular}
}
\end{center}
\caption{F5-TTS Base model evaluation results on LibriSpeech-PC \textit{test-clean}, LibriSpeech-non-PC \textit{test-clean}, Seed-TTS \textit{test-en} and \textit{test-zh} with BigVGAN and Vocos, default setting as in Sec.\ref{sec:expr_results}. The WER scores in brackets indicate results leveraging the Hubert-large-based ASR model.}
\label{tab:compr_vocoder}
\end{table*}

\begin{table*}[ht]
\begin{center}
\setlength\tabcolsep{8pt}
\resizebox{0.88\linewidth}{!}{
\begin{tabular}{cc|ccc|ccc}
\toprule
\multicolumn{2}{c}{\bf Train Set} &\multicolumn{3}{c}{\textbf{LibriTTS} - 585 hours}&\multicolumn{3}{c}{\textbf{LJSpeech} - 24 hours} \\
\cdashline{1-8}\noalign{\vskip\belowrulesep}
\multicolumn{2}{c}{\bf Test Set}  &\multicolumn{3}{c}{\bf LibriSpeech-PC \textit{test-clean}}&\multicolumn{3}{c}{\bf LJSpeech in-set tests} \\
\bf Model (\#Param.) & \bf Update &{\bf { }WER(\%)↓} &{\bf SIM-o ↑} &{\bf UTMOS ↑{ }} &{\bf { }WER(\%)↓} &{\bf SIM-o ↑} &{\bf UTMOS ↑{ }} \\
\midrule
Ground Truth &-    &{ }2.23  &0.69  &4.09{ }    &{ }2.36  &0.72  &4.36{ } \\
USLM (361M)  &-    &{ }6.1   &0.43  &-{ }       &{ }-     &-     &-{ }    \\
\cdashline{1-8}\noalign{\vskip\belowrulesep}
             &100K &{ }29.5  &0.53  &3.78{ }    &{ }5.64  &0.72  &4.17{ } \\
             &200K &{ }4.58  &0.59  &4.07{ }    &{ }2.93  &0.72  &4.18{ } \\
F5-TTS small &300K &{ }2.71  &0.60  &4.11{ }    &{ }3.26  &0.71  &4.12{ } \\
(158M)       &400K &{ }2.44  &0.60  &4.11{ }    &{ }3.90  &0.70  &4.05{ } \\
             &500K &{ }2.20  &0.60  &4.10{ }    &{ }4.68  &0.70  &3.99{ } \\
             &600K &{ }2.23  &0.59  &4.10{ }    &{ }5.25  &0.69  &3.93{ } \\
\bottomrule
\end{tabular}
}
\end{center}
\caption{F5-TTS small models evaluation results on LibriSpeech-PC \textit{test-clean} (model trained on LibriTTS 585 hours multi-speaker dataset), and on LJSpeech in-set test samples (model trained on 24 hours single-speaker LJSpeech); Vocos as vocoder, Whisper-large-v3 as ASR model. The scores of USLM \citep{speechtokenizer} are evaluated with the official checkpoint pre-trained on LibriTTS. }
\label{tab:compr_data_scale}
\end{table*}

\subsection{Training and Inference Performance with Different Dataset Scales}
\label{appx:compr_data_scale}

We train F5-TTS 158M small models on LibriTTS \citep{libritts} 585 hours and LJSpeech \citep{ljspeech} 24 hours English datasets to provide insights on our model's training stability with different dataset scales, typically to see whether it can maintain stable training on limited data. Both training takes place with the same configuration as described in Sec.\ref{sec:modelarchitecture} and Appendix \ref{appx:smallmodels} despite a batch size of 307,200 audio frames (0.91 hours) as base models. Every 100K update takes approximately 8 hours on 8 NVIDIA H100 SXM GPUs.

Same as Sec.\ref{sec:expr_results}, we report the average score of three random seed generation results, using a CFG strength of 2, a Sway Sampling coefficient of $-1$, and 32 NFE steps. Since LJSpeech is a single-speaker dataset, we measure the metrics on in-set tests (1000 samples organized with 4 to 7 seconds to infer with an around 3-second prompt). It is clear from Tab.\ref{tab:compr_data_scale} (and Fig.\ref{fig:modelarchitecture} in comparison with E2 TTS small) that our design enables stable training to learn speech-text alignment (without grapheme-to-phoneme) with varying data amounts.

\section{Subjective Evaluation Details}
\label{appx:subj_eval}

To evaluate speech quality, we conduct a CMOS subjective evaluation. 20 natives were invited for both English and Mandarin to evaluate 30 rounds with randomly selected utterances for all three test sets and all model variants. Evaluators were informed in detail about the guidelines and scoring criteria for the CMOS test, for example, the general instruction part:

\begin{itemize}
  \item Most important: use high-quality studio headphones and a good sound card!
  \item Listen through all test files and test sets before you do any ratings to get used to the material.
  \item Rate the quality of the test items only compared to the reference on top.
  \item Try to rate the overall impression of a test item and don't concentrate on single aspects.
\end{itemize}

For the CMOS test, the overall quality of a generated speech is first rated from $-3$ (bad quality compared to the reference) to $+3$ (much better than the reference) integer scale, then reported in average differentials with received scores of ground truth speech. For SMOS, a 1 to 5 with 0.5 as an interval rating is employed (higher better). Judges are to score the similarity between the synthesized and prompt speech with clearly differentiated instructions mentioning:

\begin{itemize}
  \item Try to rate concentrating on the speaker similarity aspects with reference speech.
\end{itemize}

We encourage more rigorous and transparent subjective evaluations, such as releasing used samples if not open-sourcing the model checkpoints. Meanwhile inviting more evaluators leads to more comprehensive and fair rating scores. Just for reference, DiTTo-TTS \citep{dittotts} received and reported 6 and 12 ratings for SMOS and CMOS, respectively, NaturalSpeech 3 \citep{ns3} invited 12 natives to judge 20 samples for CMOS and 10 samples for SMOS.

\end{document}